\documentclass[12pt,a4paper]{article}

\usepackage{jheppub}

\usepackage{etoolbox}
    \makeatletter
    \patchcmd{\maketitle}{\@fpheader}{}{}{}
    \makeatother

\usepackage[utf8]{inputenc}
\usepackage{lmodern}
\usepackage{exscale}

\usepackage{amsmath}
\usepackage{amssymb}
\usepackage{mathtools}

\usepackage{mathrsfs}

\usepackage{tensor}

\usepackage[low-sup]{subdepth}

\newcommand{\scr}{\scriptscriptstyle}

\usepackage{graphicx}
\newcommand{\longsim}{\scalebox{1.8}[1]{$\sim$}}

\usepackage{xcolor}
\hypersetup{
    colorlinks,
    linkcolor={black},
    citecolor={black},
    urlcolor={black}
}

\usepackage[normalem]{ulem}

\usepackage{cancel}

\title{Photon propagator in de Sitter space \\ in the general covariant gauge}

\author[a]{Dra\v{z}en Glavan,}
\emailAdd{glavan@fzu.cz}
\author[b]{Tomislav Prokopec}
\emailAdd{t.prokopec@uu.nl}

\affiliation[a]{CEICO, Institute of Physics of the Czech Academy of Sciences,
	\\
	Na Slovance 1999/2, 182 21 Prague 8, Czech Republic}
\affiliation[b]{Institute for Theoretical Physics, Spinoza Institute \& EMME$\Phi$,
	\\
	Utrecht University, Buys Ballot Building, Princetonplein 5,
	\\
	3584 CC Utrecht, The Netherlands}

\abstract{
We consider a free photon field in~$D$-dimensional de Sitter space, and construct its 
propagator in the general covariant gauge. Canonical quantization is employed to define the
system starting from the classical theory.
This guarantees that the propagator satisfies both the equation of motion~{\it and} subsidiary
conditions descending from gauge invariance and gauge fixing.
We first construct the propagator as a sum-over-modes in momentum space, 
carefully accounting for symmetry properties of the state.
We then derive the position space propagator in a covariant representation, that is our main result.
Our conclusions disagree with previous results
as we find that the position space photon propagator {\it necessarily breaks de Sitter symmetry}, except
in the exact transverse gauge limit.

}

\notoc

\begin{document}

\maketitle

\titlepage

\section{Introduction}
\label{sec: Introduction}

Linear dynamics of a massless vector field --- the photon --- in expanding cosmological spaces
is considered to be
 particularly simple and well understood. The photon couples conformally
to gravity and thus effectively does not sense the expansion. 
This makes the dynamics of its two physical polarizations no more complicated than 
in flat space.
However, the conformal coupling of the photon
can be broken via couplings to non-trivial field condensates or other
non-conformally coupled fields such as scalars or gravitons. 
Primordial inflation is where these effects can be particularly important because
of the huge scale of the expansion rate.
There are a number of cases where vector fields play an important role in inflation,
some of which are vector 
inflation~\cite{Dimopoulos:2006ms,Golovnev:2008cf,Dimopoulos:2008rf,BeltranJimenez:2008zzi,Dimopoulos:2009vu,Maleknejad:2012fw},
preheating after inflation~\cite{Garcia-Bellido:2008ycs,Ema:2016dny},
axion inflation~\cite{Adshead:2016iae,Adshead:2015pva,Pajer:2013fsa,Barnaby:2011qe},
 inflationary magnetogenesis
from tree level dynamics~\cite{Turner:1987bw,Ratra:1991bn,Prokopec:2001nc}
(for a review see~\cite{Giovannini:2003yn})
and from dynamically generated 
condensates during inflation~\cite{Davis:2000zp,Dimopoulos:2001wx},
the Schwinger effect in 
inflation~\cite{Garriga:1994bm,Martin:2007bw,Kobayashi:2014zza,Frob:2014zka,Lozanov:2018kpk,Banyeres:2018aax,Rajeev:2019okd,Domcke:2021fee}, and 
corrections to the standard model and inflaton effective 
potential~\cite{Garbrecht:2006aw,George:2012xs,Miao:2015oba,Markkanen:2018bfx,Liao:2018sci,Miao:2019bnq,Katuwal:2021kry,Katuwal:2022szw}.
Particularly interesting are cases where conformal coupling is broken by 
quantum loop effects in inflation. 
This happens for interactions with light
spectator scalar fields~\cite{Prokopec:2002jn,Prokopec:2002uw,Prokopec:2003bx,Prokopec:2003iu,Kahya:2005kj,Kahya:2006ui,Prokopec:2006ue,Prokopec:2007ak,Prokopec:2008gw,Leonard:2012si,Leonard:2012ex,Chen:2016nrs,Chen:2016uwp,Chen:2016hrz,Kaya:2018qbj,Popov:2017xut,Glavan:2019uni}
or with inflationary gravitons~\cite{Leonard:2012fs,Leonard:2013xsa,Glavan:2013jca,Wang:2014tza,Glavan:2015ura,Wang:2015eaa,Glavan:2016bvp,Miao:2018bol}. Both cases point to large non-perturbative effects.
A  useful idealization
of the slow-roll inflationary spacetime is a rigid de Sitter space, which is one of the 
three maximally symmetric spacetimes, and maximal symmetry implies considerable 
conceptual and computational simplifications. 
In this work we consider~$D$-dimensional photon two-point functions in the expanding patch
of de Sitter space, appropriate for dimensionally regulated, nonequilibrium
perturbative computations in the Schwinger-Keldysh (also known as {\it in-in} or 
closed-time-path) formalism.

While at the linear level we have the luxury of explicitly isolating the physical
degrees of freedom --- the two transverse polarizations --- this is no longer as 
straightforward at the interacting level. 
Instead, it is preferable and considerably simpler
to perform computations in a particular gauge with gauge-dependent quantities,
and only later project out the physical information.
Fixing the gauge can be 
done in a number of ways, but the choice preferred for computations is the one where no
field components are explicitly eliminated. These are referred
to as average gauges (or multiplier gauges), and
are characterized by adding a gauge-fixing term to the original 
gauge-invariant action. Arguably the most natural
choice for the gauge-fixing term is the
{\it general covariant gauge,}
\begin{equation}
S_{\rm gf}[A_\mu] = \int\! d^{D\!}x \, \sqrt{-g} \, \biggl[
	- \frac{1}{2\xi} \bigl( g^{\mu\nu} \nabla_\mu A_\nu \bigr)^{\!2} \,
	\biggr] \, ,
\label{intro gauge}
\end{equation}
that comes with one free real parameter~$\xi$, defining a one-parameter
family of gauges. Any physical quantity must be independent on $\xi$, which makes these gauges 
particularly useful.
It is commonly held that photon propagators in the general covariant gauge are
de Sitter invariant, which is supported by the existing literature. It his work we challenge
this belief.

Covariant gauge propagators for
different values of the gauge-fixing parameters have been derived in several works,
starting with the seminal work by Allen and Jacobson~\cite{Allen:1985wd}. Among several 
different cases and spacetimes, they derived the photon propagator in de Sitter 
space of arbitrary spacetime dimension~$D$, in~$\xi\!=\!1$ 
covariant gauge. To this end they made a de Sitter invariant ansatz for the propagator,
solved the resulting simplified equations of motion, and fixed the ambiguities by considering the 
singularity structure. This method was used in several subsequent works to derive propagators
in different covariant gauges. In~\cite{Kahya:2005kj} 
the propagator from~\cite{Allen:1985wd} was rederived in a somewhat different form.
Tsamis and Woodard~\cite{Tsamis:2006gj} used the method when considering 
a massive vector propagator whose massless limit corresponds to the  photon
in Landau gauge~$\xi\!\to\!0$ in~$D$-dimensional de Sitter. {To obtain the 
propagator they required  their {\it Ansatz} to be transverse, in addition to being de Sitter invariant.
The methods from~\cite{Allen:1985wd} was also used by Youssef~\cite{Youssef:2010dw}
to derive the photon propagator for arbitrary gauge-fixing parameter in four space-time dimensions.
The most general photon propagator, valid for arbitrary~$\xi$ and~$D$,
was reported by Fr\"ob and Higuchi~\cite{Frob:2013qsa}. Unlike the preceding works,
they utilized canonical quantization of the gauge-fixed vector sector of the Stueckelberg
model, containing a massive vector field. The resulting propagator they report was derived
as a sum-over-modes, and its de Sitter invariant massless limit encompasses all previous results
as special cases.~\footnote{In Ref.~\cite{Domazet:2014bqa} the method of making the de Sitter invariant
{\it Ansatz} was considered for the case of general~$\xi$ and~$D$. The analysis there would have
reproduced the de Sitter invariant result of~\cite{Frob:2013qsa}, had the integrals in the 
result reported been evaluated.}

It would seem that little more could be said about covariant gauge photon propagators in de 
Sitter. Nevertheless, it was pointed out recently~\cite{Glavan:2022a} that photon propagators in so-called
average (or multiplier) gauges, in addition to solving the equations of motion,
must satisfy subsidiary conditions that are a consequence of gauge symmetry. 
These subsidiary conditions derive from the quantization of the first-class constraints 
of the classical theory, and amount to the condition that correlators of first-class constraints
have to vanish in the quantized theory. 
It was found that the photon propagators reported in the literature do not satisfy all the 
subsidiary conditions~\cite{Glavan:2022a}, except in the gauge~$\xi\!\to\!0$.
This is a problem that needs to be addressed, and
it motivates us here to consider the construction of the photon propagator in covariant gauges
from the first principles of canonical quantization.

The canonical quantization is based on the canonical structure only, and is conceptually divorced
from the symmetries of the background spacetime. Even though making this distinction is
often not necessary, here we find it important when trying to understand where the problem
with the propagators comes from and how to resolve it.
We pay special attention to how particular symmetry properties of quantum states are imposed,
by considering conserved charges from both the gauge-invariant and gauge-fixed formulations,
that serve as generators of de Sitter symmetry transformations. Our main result is
rather surprising --- the physically de Sitter invariant quantum state of the photon does not
admit a de Sitter invariant propagator in the general covariant gauge (except in the 
limit~$\xi\!\to\!0$). We derive this result by representing
the propagator as a sum-over-modes in momentum space, 
which we subsequently solve to find the propagator in position space.
The resulting expression consists of a de Sitter invariant part, that corresponds to the massless
limit of~\cite{Frob:2013qsa}, and a previously missed de Sitter breaking part which ensures that
the subsidiary conditions are respected. In a companion letter~\cite{GlavanProkopec:2022}
we show how one can obtain the same result in position space,
starting from BRST quantization and utilizing Ward-Takahashi identities.


This paper is organized in eight sections, the first of which is concluding. 
The following section collects definitions and results on propagators and scalar mode
functions used throughout the paper.
The third section recounts the canonical formulation of the photon
in multiplier gauges and canonical quantization.
The dynamics of field operators is solved for in the fourth section, and
the fifth section is devoted to the construction of the quantum state and discussion of
its symmetries. The main results are derived in the sixth section, that
gives the solution for the propagator satisfying all the required subsidiary conditions.
Checks of the propagator solution are performed in the seventh section
by considering two simple observables, while the concluding eighth section 
contains a discussion of the main results. More technical details are relegated 
to two appendices.

\section{Preliminaries}
\label{sec: Preliminaries}

This section collects definitions and results frequently used in subsequent sections.
First the Poincar\'{e} patch of de Sitter space is defined, and then some useful results
on scalar two-point functions and scalar mode functions in de Sitter are recalled.

\subsection{De Sitter space}
\label{subsec: De Sitter space}

The invariant line element of the $D$-dimensional Friedman-Lema\^itre-Robertson-Walker (FLRW) spacetime,
\begin{equation}
ds^2 
	= - dt^2 + a^2(t) d\vec{x}^{\,2}
	= a^2(\eta) \bigl[ -d\eta^2  +d\vec{x}^{\,2} \bigr] \, ,
\end{equation}
defines the associated conformally flat metric~$g_{\mu\nu}\!=\! a^2(\eta) {\rm diag} (-1 , 1 , \dots , 1)$,
and where the speed of light is taken to be unity,~$c\!=\!1$.
Here Cartesian coordinates~$\vec{x}$ span~$(D\!-\!1)-$dimensional flat
spatial slices, while time is parametrized
either by the physical time~$t$, or by conformal time~$\eta$. The two
times are related by~$dt\!=\!a(\eta) d\eta$, where~$a$ is the scale factor that encodes the 
dynamics of the expansion, which is usually expressed either in terms of the physical Hubble 
rate,~$H \!=\! (da/dt)/a$, or the conformal Hubble rate,~$\mathcal{H}\!=\! (da/d\eta) / a$,
the two being related by~$\mathcal{H} \!=\! aH$.

The expanding Poincar\'{e} patch of de Sitter space is defined as the FLRW spacetime with
a constant physical Hubble rate,~$H = {\tt const.}$, when the conformal Hubble rate 
and the scale factor take the following functional form,
\begin{equation}
\mathcal{H} = \frac{H}{1 \!-\! H(\eta\!-\!\eta_0)} \, ,
\qquad \qquad
a(\eta) = \frac{\mathcal{H}}{H} \, ,
\end{equation}
where~$\eta_0$ is the initial time, for which~$a(\eta_0)\!=\!1$.
The conformal time ranges on the interval~$\eta\!\in\!(-\infty,\eta_0 \!+\! 1/H)$.

\subsection{Scalar two-point functions}
\label{Scalar two-point functions}

Two-point functions of scalar fields often appear as building blocks of two-point functions for 
higher spin fields in de Sitter. This will be true for the photon propagator we construct in this work,
so here we summarize and recall some of the properties of scalar field mode functions and the two-point
functions constructed out of them.

The positive-frequency Wightman function~\footnote{Our naming for two-point functions
follows the Keldysh polarity conventions.}
can be taken as the elementary two-point function.
It satisfies a homogeneous equation of motion,
\begin{equation}
\bigl( \, \square - M_\lambda^2 \, \bigr) i \bigl[ \tensor*[^{ \! \scr -\!}]{\Delta}{^{\scr \!+ \!} } \bigr]_\lambda(x;x') = 0 \, ,
\label{scalar EOM: Wightman}
\end{equation}
where~$\square \!=\! g^{\mu\nu} \nabla_\mu \nabla_\nu$ is the d'Alembertian, and where the effective mass,
conveniently parametrized by~$\lambda$,
\begin{equation}
M_\lambda^2 = \biggl[ \Bigl( \frac{D\!-\!1}{2} \Bigr)^{\!2} - \lambda^2 \biggr] H^2
\,,
\label{Mass parameter lambda}
\end{equation}
receives contributions from both the scalar field mass and its non-minimal coupling to the Ricci scalar.
The negative frequency Wightman function is just a complex 
conjugate of the positive-frequency 
one,~$ i \bigl[ \tensor*[^{ \! \scr +\!}]{\Delta}{^{\scr \!- \!} } \bigr]_\lambda(x;x')  \!=\! 
	\bigl\{  i \bigl[ \tensor*[^{ \! \scr -\!}]{\Delta}{^{\scr \!+ \!} } \bigr]_\lambda(x;x') \bigr\}^*$,
and satisfies the same equation of motion.
The Feynman propagator 
involves time-ordering, and is expressed in terms of the Heaviside step function as,
\begin{equation}
i \bigl[ \tensor*[^{\scr \!+\!}]{\Delta}{^{\scr \!+\! } } \bigr](x;x') = 
	\theta(\eta \!-\! \eta') i \bigl[ \tensor*[^{\scr \!-\!}]{\Delta}{^{\scr \!+\! } } \bigr](x;x') 
	+
	\theta(\eta' \!-\! \eta) i \bigl[ \tensor*[^{\scr \!+\!}]{\Delta}{^{\scr \!-\! } } \bigr](x;x') 
	\, .
\label{scalar Feynman def}
\end{equation}
and satisfies a sourced equation of motion,
\begin{equation}
\bigl( \, \square - M_\lambda^2 \, \bigr)  i \bigl[ \tensor*[^{ \! \scr +\!}]{\Delta}{^{\scr \!+ \!} } \bigr]_\lambda(x;x') 
	=	\frac{ i \delta^{D}(x \!-\! x') }{ \sqrt{-g} }
\, .
\label{scalar Feynman eq}
\end{equation}
The Dyson propagator is its complex 
conjugate,~$i \bigl[ \tensor*[^{\scr \!-\!}]{\Delta}{^{\scr \!-\! } } \bigr](x;x') 
	\!=\! \bigl\{ i \bigl[ \tensor*[^{\scr \!+\!}]{\Delta}{^{\scr \!+\! } } \bigr](x;x') \bigr\}^* $,
and satisfies a conjugate of Eq.~(\ref{scalar Feynman eq}).

The scalar two-point functions admit a sum-over-modes representation,
\begin{equation}
i \bigl[ \tensor*[^{\scr \! - \!}]{\Delta}{^{\scr \! +\! }} \bigr]_\lambda(x;x')
	=
(aa')^{-\frac{D-2}{2}} \!\!
\int\! \frac{d^{D-1}k }{ (2\pi)^{D-1 } } \, e^{i\vec{k} \cdot (\vec{x} - \vec{x}^{\, \prime}) } \,
	\mathscr{U}_\lambda(\eta,\vec k) \bigl[ \mathscr{U}_\lambda(\eta',\vec k) \bigr]^*
	\, ,
\label{int over modes: Wightman}
\end{equation}
in terms of the conformally rescaled scalar field mode function~$\mathscr{U}_\lambda(\eta,k)$. 
We summarize the 
properties of the scalar mode functions in the following subsection.
The integral in Eq.~(\ref{int over modes: Wightman}) is generally divergent for real coordinates,
and analytic continuation is called for. The prescription
$\eta\!\to\! \eta \!-\! i\varepsilon/2$ and $\eta' \!\to\! \eta' \!-\! i \varepsilon/2$ 
preserves its naive properties under complex conjugation, and
defines it as a distributional limit~$\varepsilon\!\to\! 0_{\scr +}$.
Depending on the value of parameter~$\lambda$ solutions
for two-point functions are qualitatively different.
For~$\lambda\!<\! (D\!-\!1)/2$ there are de Sitter invariant solutions, while 
for~$\lambda\!\ge\! (D\!-\!1)/2$ they do not exist,
 see {\it e.g.} Ref.~\cite{Janssen:2008px}. 
We summarize these two cases for position space
scalar two-point functions in the two concluding subsections.

\subsubsection{Scalar mode functions}
\label{subsec: Scalar mode functions}

The equation of motion for scalar mode functions in de Sitter is,
\begin{equation}
\biggl[ \partial_0^2 + k^2 - \Bigl( \lambda^2 \!-\! \frac{1}{4} \Bigr) \mathcal{H}^2 \biggr]
	\mathscr{U}_{\lambda}(\eta,\vec{k} ) 
	= 0 \, ,
\label{mode eq}
\end{equation}
where~$\lambda$ is a constant defined in~(\ref{Mass parameter lambda}).
Its general solution in power-law inflation is,
\begin{equation}
\mathscr{U}(\eta, \vec{k}) 
	= \alpha(\vec{k}) \, U_\lambda(\eta,k)
	+ \beta(\vec{k}) \, U_\lambda^*(\eta,k) \, ,
\label{general mode solution}
\end{equation}
with the positive-frequency Chernikov-Tagirov-Bunch-Davies (CTBD) mode 
function~\cite{Chernikov:1968zm,Bunch:1978yq},
\begin{equation}
U_\lambda(\eta,k) = 
	e^{\frac{i\pi}{4} (2\lambda+1) } e^{ \frac{- i k}{ H } } 
	\sqrt{ \frac{ \pi }{ 4 \mathcal{H} } } \,
	H_\lambda^{\scr (1)}\Bigl( \frac{ k }{ \mathcal{H} } \Bigr)
	\, .
\label{CTBD solution}
\end{equation}
where~$H_\lambda^{\scr (1)}$ is the Hankel function of the first kind,
and $\alpha(\vec k)$ and $\beta(\vec k)$ are complex 
Bogolybov coefficients satisfying, 
$|\alpha(\vec k)|^2-|\beta(\vec k)|^2=1$.
The flat space limit of the this mode function is
\begin{equation}
U_\lambda(\eta,k)
	\, \overset{H\to0}{\longsim} \,
	\frac{ e^{- i k (\eta-\eta_0)} }{ \sqrt{2k} }
	\biggl\{
	1 + \Bigl( \lambda^2 \!-\! \frac{1}{4} \Bigr) 
		\biggl[ \frac{ iH }{ 2k }
			+
	\Bigl( \lambda^2 \!-\! \frac{9}{4} - 4 i k (\eta\!-\!\eta_0) \Bigr)
		\frac{ H^2 }{ 8k^2 }
		+ \mathcal{O}(H^3) \biggr]
	\biggr\}
	\, ,
\label{CTBD def T}
\end{equation}
where~$\eta_0$ is the initial time at which~$a(\eta_0)\!=\!1$.
We make a frequent use of recurrence relations between contiguous scalar mode functions,
\begin{align}
\biggl[ \partial_0 + \Bigl( \lambda \!+\! \frac{1}{2} \Bigr) \mathcal{H} \biggr] U_\lambda
	= -ik U_{\lambda+1} \, ,
\qquad 
\biggl[ \partial_0 - \Bigl( \lambda \!+\! \frac{1}{2} \Bigr) \mathcal{H} \biggr] U_{\lambda+1}
	= -ik U_{\lambda} \, ,
\label{mode recurrence}
\end{align}
which follow from the recurrence relations for Hankel functions~({\it c.f.}~(10.6.2)
 in~\cite{Olver:2010}).
Using these the Wronskian is conveniently written as,
\begin{equation}
{\rm Re} \Bigl[ U_\lambda(\eta,k) U_{\lambda+1}^*(\eta,k) \Bigr] = \frac{1}{2k} \, . 
\label{Wronskian condition}
\end{equation}
We also use two identities following from 
the equation of motion~(\ref{mode eq}), 
\begin{align}
\biggl[ \partial_0^2 + k^2 
	- \Bigl( \lambda^2 \!-\! \frac{1}{4} \Bigr) 
		\mathcal{H}^2 \biggr] \Bigl( \mathcal{H} U_{\lambda+1} \Bigr)
		={}&
		2 \mathcal{H}^3 \biggl[
		2 ( \lambda \!+\! 1 ) U_{\lambda+1}
		- \frac{ik}{\mathcal{H}} U_\lambda
		\biggr]
		\, ,
\label{mode id1}
\\
\biggl[ \partial_0^2 + k^2 
	- \Bigl( \lambda^2 \!-\! \frac{1}{4} \Bigr) 
		\mathcal{H}^2 \biggr] \frac{\partial U_\lambda}{\partial \lambda}
		={}&
		2 \lambda \mathcal{H}^2 U_\lambda
		\, ,
\label{mode id2}
\end{align}
the former one by applying the recurrence relations~(\ref{mode recurrence}),
and the later by taking a parametric derivative of~(\ref{mode eq}).

\subsubsection{De Sitter invariant scalar two-point functions}
\label{subsubsec: De Sitter invariant scalar propagators}

For~$\lambda\!<\! (D\!-\!1)/2$,
the positive-frequency Wightman function of the real scalar field~(\ref{scalar EOM: Wightman})
 in power-law inflation
has a sum-over-modes representation~(\ref{int over modes: Wightman}),
\begin{equation}
i \bigl[ \tensor*[^{\scr \! - \!}]{\Delta}{^{\scr \! +\! }} \bigr]_\lambda(x;x')
	=
(aa')^{-\frac{D-2}{2}} \!\!
\int\! \frac{ d^{D-1}k }{ (2\pi)^{D-1 } } \, e^{i\vec{k} \cdot (\vec{x} - \vec{x}^{\, \prime}) } \,
	U_\lambda(\eta,k) U_\lambda^*(\eta',k) .
\label{int over modes}
\end{equation}
This integral representation evaluates to,
\begin{equation}
i \bigl[ \tensor*[^{\scr \! - \!}]{\Delta}{^{\scr \! + \!}} \bigr]_\lambda(x;x')
	=
	\mathcal{F}_\lambda(y_{\scr -+}) \, ,
\label{scalar prop solution}
\end{equation}
where~$y_{\scr -+}$ is the~$i\varepsilon$-regulated distance function appropriate for 
the positive-frequency Wightman function,
\begin{equation}
y_{\scr_{-+}} (x;x')
	= \mathcal{H} \mathcal{H}' \bigl( \Delta x^2_{\scr -+} \bigr)
	= \mathcal{H} \mathcal{H}'
		\Bigl[ \| \vec{x} \!-\! \vec{x}^{\,\prime} \|^2 - \bigl( \eta \!-\! \eta' \!-\! i \varepsilon \bigr)^2 \Bigr] \, ,
\label{y Wightman}
\end{equation}
and the propagator function is
expressed in terms of a hypergeometric function,
\begin{align}
\mathcal{F}_\lambda(y)
	={}&
	\frac{ H^{D-2} }{ (4\pi)^{ \frac{D}{2} } }
	\frac{ \Gamma\bigl( \frac{D-1}{2} \!+\! \lambda \bigr) \, 
		\Gamma\bigl( \frac{D-1}{2} \!-\! \lambda \bigr) }{ \Gamma\bigl( \frac{D}{2} \bigr) }
\nonumber \\
&	\hspace{2cm}
	\times
	{}_2F_1\biggl( \Bigl\{ \frac{D\!-\!1}{2} \!+\! \lambda , \frac{D\!-\!1}{2} \!-\! \lambda \Bigr\} , 
		\Bigl\{ \frac{D}{2} \Bigr\} , 1 \!-\! \frac{y}{4} \biggr)
		\, ,
\label{F def}
\end{align}
that satisfies the equation the hypergeometric equation in a different guise,
\begin{equation}
\biggl[
	(4y \!-\! y^2) \frac{\partial^2}{\partial y^2}
	+ D (2\!-\!y) \frac{ \partial }{ \partial y }
	+ \lambda^2 - \Bigl( \frac{ D\!-\!1 }{2} \Bigr)^{\!2} \,
	\biggr]
	\mathcal{F}_\lambda(y) 
	= 0
	\, .
\label{F eom}
\end{equation}
The scalar Feynman propagator~(\ref{scalar Feynman def}) 
-- known as the Chernikov-Tagirov propagator~\cite{Chernikov:1968zm} --
takes the same form as the Wightman function,
\begin{equation}
i \bigl[ \tensor*[^{\scr \! + \!}]{\Delta}{^{\scr \! + \! } } \bigr]_\lambda(x;x')
	=
	\mathcal{F}_\lambda(y_{\scr ++}) 
	\, ,
\end{equation}
where the argument of the propagator function is substituted by the appropriate one
according to~(\ref{scalar Feynman def}),~\footnote{
 The~$i\varepsilon$ prescription for the Feynman propagator 
in~(\ref{i varpsilon prescription: Feynman})
follows from the prescription for the Wightman function in~(\ref{y Wightman})
and the definition~(\ref{scalar Feynman def}), upon using the properties of the step 
function: $\theta(\Delta\eta) \!+\! \theta(-\Delta\eta)\!=\!1$,~$[\theta(\Delta\eta)]^2\!=\! \theta(\Delta\eta)$, 
and~$\theta(\Delta\eta) \!\times \!\theta(-\Delta\eta) \!=\! 0$, where~$\Delta\eta\!=\! \eta\!-\! \eta'$.
}
}
\begin{equation}
y_{\scr ++} 
	=  \mathcal{H} \mathcal{H}' \bigl( \Delta x^2_{\scr ++} \bigr)
	= \mathcal{H} \mathcal{H}'
		\Bigl[ \| \vec{x} \!-\! \vec{x}^{\,\prime} \|^2 - \bigl( | \eta \!-\! \eta' | \!-\! i \varepsilon \bigr)^2 \Bigr] \, .
\label{i varpsilon prescription: Feynman}
\end{equation}
Henceforth we suppress the polarity indices denoting different
$i\varepsilon$-prescriptions for distance functions~$y$ that distinguish between
two-point functions, as they should be clear from the context.

A useful representation of the function in~(\ref{F def}) is a power series around~$y\!=\!0$,
\begin{align}
\MoveEqLeft[1]
\mathcal{F}_\lambda(y) 
	=
	\frac{ H^{D-2} \, \Gamma\bigl( \frac{D-2}{2} \bigr) }{ (4\pi)^{ \frac{D}{2} } } 
	\biggl\{
	\Bigl( \frac{y}{4} \Bigr)^{\! -\frac{D-2}{2} }
	+ \frac{ \Gamma\bigl( \frac{4-D}{2} \bigr) }{ \Gamma\bigl( \frac{1}{2} \!+\! \lambda \bigr) \, \Gamma\bigl( \frac{1}{2} \!-\! \lambda \bigr) } 
		\sum_{n=0}^{\infty}
\label{power series}
\\
&
	\times \!
	\biggl[
	\frac{ \Gamma\bigl( \frac{3}{2} \!+\! \lambda \!+\! n \bigr) \, \Gamma\bigl( \frac{3}{2} \!-\! \lambda \!+\! n \bigr) }
		{ \Gamma\bigl( \frac{6-D}{2} \!+\! n \bigr) \, (n\!+\!1)! } \Bigl( \frac{y}{4} \Bigr)^{\!n - \frac{D-4}{2}}
	-
	\frac{ \Gamma\bigl( \frac{D-1}{2} \!+\! \lambda \!+\! n \bigr) \, \Gamma\bigl( \frac{D-1}{2} \!-\! \lambda \!+\! n \bigr) }
		{ \Gamma\bigl( \frac{D}{2} \!+\! n \bigr) \, n! }
	\Bigl( \frac{y}{4} \Bigr)^{\!n }
	\biggr]
	\biggr\} \, .
\nonumber 
\end{align}
Furthermore, 
Gauss' relations between hypergeometric functions ({\it cf.} (9.137) of~\cite{Gradshteyn:2007})
allow us to derive recurrence relations between contiguous propagator functions,
\begin{align}
2 \frac{\partial \mathcal{F}_\lambda}{\partial y}
	={}&
	(2\!-\!y) \frac{\partial \mathcal{F}_{\lambda+1} }{ \partial y }
	+ \Bigl( \lambda \!-\! \frac{D\!-\!3}{2} \Bigr) \mathcal{F}_{\lambda+1}
	\, ,
\label{contiguousF 1}
\\
2 \frac{\partial \mathcal{F}_{\lambda+1}}{\partial y}
	={}&
	(2\!-\!y) \frac{\partial \mathcal{F}_{\lambda} }{ \partial y }
	- \Bigl( \lambda \!+\! \frac{D\!-\!1}{2} \Bigr) \mathcal{F}_{\lambda}
	\, ,
\label{contiguousF 2}
\end{align}
that we utilize in Sec.~\ref{sec: Two-point function}.

\subsubsection{De Sitter breaking scalar two-point functions}
\label{subsubsec: De Sitter breaking scalar propagators}

When the index of the scalar mode function is~$\lambda \!>\! (D\!-\!1)/2$, the CTBD mode 
function leads to an unphysical infrared divergent Wightman function. In such cases the 
physical mode function must be modified in the infrared by introducing Bogolyubov
coefficients in~(\ref{general mode solution}) that suppress the singular behaviour,
and lead to a well defined sum-over-modes
representation~(\ref{int over modes}).
Choosing them to preserve cosmological symmetries leads to the following Wightman 
function~\cite{Janssen:2008px},
\begin{equation}
i \bigl[ \tensor*[^{\scr \!-\! }]{\Delta}{^{\scr \!+\!} } \bigr]_\lambda(x;x')
	=
	\mathcal{F}_\lambda(y)
	+ \mathcal{W}_\lambda(u) 
	\, ,
\label{Wightman function with sym breaking}
\end{equation}
composed of the de Sitter invariant part~(\ref{F def}),
and the de Sitter breaking part,
\begin{equation}
\mathcal{W}_\lambda(u) 
	=
	\frac{H^{D-2}}{ (4\pi)^{\frac{D}{2} } }
	\frac{ \Gamma(2\lambda) \, \Gamma(\lambda) }
		{ \Gamma\bigl( \frac{D - 1}{2} \bigr) \, \Gamma\bigl( \frac{1}{2} \!+\! \lambda \bigr) }
		\frac{ e^{ ( \lambda - \frac{D-1}{2} ) u } }{ \lambda \!-\! \frac{D-1}{2} }
		\Bigl( \frac{k_0}{H} \Bigr)^{\! D-1-2\lambda}
		\, ,
\label{W lambda}
\end{equation}
where~$k_0$ is some infrared scale, and,
\begin{equation}
u=\ln(aa'),
\end{equation}
is a convenient bi-local scalar variable.
The Feynman propagator is then inferred from the Wightman function
in the same way as described in Sec.~\ref{subsubsec: De Sitter invariant scalar propagators}, 
by changing the implicit~$i\varepsilon$-prescription of~$y$.

For our purposes the limiting
case of the massless, minimally coupled scalar,~$\lambda\!\to\!(D\!-\!1)/2$,
is particularly important. In this limit the two-point function~(\ref{Wightman function with sym breaking})
reproduces the finite Onemli-Woodard two-point function~\cite{Onemli:2002hr},
as the divergence in the 
de Sitter breaking part cancels the one from the de Sitter invariant part, which is divergent in
any number of dimensions. Nevertheless, the quantity~$\partial \mathcal{F}_\lambda/\partial y$
is finite if first the derivative is performed, and then the limit~$\lambda\!\to\!(D\!-\!1)/2$ is taken.
Another important expression that we encounter is,
\begin{equation}
\Bigl( \frac{D\!-\!1}{2} \!-\! \lambda \Bigr) \mathcal{F}_\lambda(y)
	\xrightarrow{ \lambda\to \frac{D-1}{2} }
	\frac{ H^{D-2} }{ (4\pi)^{\frac{D}{2}} }
		\frac{ \Gamma(D\!-\!1) }{ \Gamma\bigl( \frac{D}{2} \bigr) }
		\, ,
\label{limit of F nu+1}
\end{equation}
which is valid for arbitrary $y$, and
which allows us to use recurrence relations~(\ref{contiguousF 1}) and~(\ref{contiguousF 2})
in this limit.

\section{Photon in FLRW}
\label{sec: Photon in FLRW}

The free photon in~$D$-dimensional curved space is defined by the covariant action,
\begin{equation}
S[A_\mu] = \int\! d^{D\!}x \, \sqrt{-g} \, \biggl[ - \frac{1}{4} g^{\mu\rho} g^{\nu\sigma} F_{\mu\nu} F_{\rho\sigma} \biggr] \, ,
\label{action}
\end{equation}
where~$F_{\mu\nu} \!=\! \partial_\mu A_\nu \!-\! \partial_\nu A_\mu$ is the vector field strength.
The action is invariant under~$U(1)$ gauge transformations,~$A_\mu(x) \!\to\! A_\mu(x) \!+\! \partial_\mu \Lambda(x)$
for some arbitrary function~$\Lambda(x)$.
Quantization of this free theory allows one to work out the two-point functions necessary
for perturbative loop computations in interacting theories containing massless
vector fields.

Canonical quantization of the photon, where all the components of the vector
potential are treated on equal footing, is based on the canonical formulation 
in the so-called {\it multiplier gauges} (also known as {\it average gauges} 
and sometimes  referred to as covariant gauges).
We begin by consider how the generally covariant  gauge~(\ref{intro gauge}) is implemented
in the classical theory. This serves to transparently
define canonical quantization, which is summarized in the concluding part of this section.

\subsection{Gauge-invariant photon}
\label{subsec: Gauge-invariant photon}

Our starting point is the canonical formulation of the gauge-invariant system~(\ref{action}),
in which we decompose the indices into spatial and temporal ones, and plug in the de Sitter metric,
\begin{equation}
S[A_\mu] = \int\! d^{D\!}x \, a^{D-4} \biggl[ \frac{1}{2} F_{0i} F_{0i} - \frac{1}{4} F_{ij} F_{ij} \biggr] 
\, .
\label{gauge invariant action}
\end{equation}
Henceforth in all expressions with decomposed indices we write them lowered, and adopt the convention
that all the repeated spatial indices are summed over. There is little advantage in trying to
maintain some sense of manifest covariance in the canonical formulation when the time
direction plays a preferred role.

The promotion of the first time derivatives of vector field components to independent
fields,~$F_{0i} \!\to\! V_i$,~$\partial_0 A_0 \!\to\! V_0$, and the introduction
of the accompanying Lagrange multipliers,~$\Pi_0$ and~$\Pi_i$ that ensure on-shell equivalence,
defines an intermediate first-order action,
\begin{align}
\mathcal{S}\bigl[ A_0, V_0, \Pi_0, A_i, V_i, \Pi_i \bigr]
	={}& 
	\int\! d^{D\!}x \, \biggl\{
		a^{D-4} \biggl[ \frac{1}{2} V_i V_i - \frac{1}{4} F_{ij} F_{ij} \biggr]
	+ \Pi_i \bigl( F_{0i} - V_i \bigr)
\nonumber \\
&	\hspace{1.4cm}
		+ \Pi_0 \bigl( \partial_0 A_0 - V_0 \bigr)
		\biggr\}
		\, ,
\label{extended action}
\end{align}
sometimes referred to as the extended action.
Solving for as many velocity fields as possible on-shell, which in this case means the spatial 
components,~\footnote{ We use the Dirac notation $\approx$ to denote weak
(on-shell) equalities that are valid at the level of equations of motion, as opposed
to~$=$
denoting strong (off-shell)
equalities that are valid at the level of the action.}
\begin{equation}
V_i \approx \overline{V}_i = a^{4-D} \Pi_i \, ,
\end{equation}
and plugging the solutions back into the action~(\ref{extended action}) then defines the canonical action,
\begin{align}
\mathscr{S} \bigl[ A_0, \Pi_0, A_i, \Pi_i, \ell \, \bigr]
	\equiv{}&
	\mathcal{S}\bigl[ A_0, V_0 \!\to\! \ell, \Pi_0, A_i, \overline{V}_i, \Pi_i \bigr]
\nonumber \\
	=& \int\! d^{D\!}x \, \Bigl[
		\Pi_0 \partial_0 A_0 + \Pi_i \partial_0 A_i - \mathscr{H} - \ell \, \Psi_1
		\Bigr] \, ,
\label{invariant canonical action}
\end{align}
where,
\begin{equation}
\mathscr{H} = \frac{ a^{4-D} }{2} \Pi_i \Pi_i
	+ \Pi_i \partial_i A_0
	+ \frac{a^{D-4}}{4} F_{ij} F_{ij} \, ,
\label{canonical Hamiltonian}
\end{equation}
is the canonical Hamiltonian density, 
and where we have relabeled~$V_0\!\to\!\ell$ to emphasize that 
in the canonical action~$\ell$ is a 
Lagrange multiplier responsible for generating the primary constraint,
\begin{equation}
\Psi_1 = \Pi_0 \, ,
\label{primary constraint}
\end{equation}
which in turn generates a secondary constraint on-shell,
\begin{equation}
\partial_0 \Psi_1 \approx \partial_i \Pi_i \equiv  \Psi_2
	\, .
\label{secondary constraint}
\end{equation}
These two form a complete set of 
 first-class constraints,~$\bigl\{ \Psi_1 , \Psi_2 \bigr\} \!=\! 0$.
The Poisson brackets of the canonical variables~\footnote{
Sometimes the canonical momenta are defined with upper indices
so as to have Poisson brackets in a seemingly covariant form. We find little use for
notation, which clearly does not reintroduce manifest covariance in the
canonical formulation, and can be misleading at times.
}
follow from the symplectic part of the canonical 
action~(\ref{invariant canonical action}),
\begin{align}
\bigl\{ A_0(\eta,\vec{x}) , \Pi_0(\eta,\vec{x}^{\,\prime}) \bigr\}
	= \delta^{D-1}(\vec{x} \!-\! \vec{x}^{\,\prime}) \, ,
\qquad
\bigl\{ A_i(\eta,\vec{x}) , \Pi_j(\eta,\vec{x}^{\,\prime}) \bigr\}
	= \delta_{ij} \delta^{D-1}(\vec{x} \!-\! \vec{x}^{\,\prime}) \, .
\label{Poisson brackets}
\end{align}
The dynamical equations descending from the canonical action do not fix the 
Lagrange multiplier~$\ell$, which can be chosen arbitrarily. This property
is the canonical formulation equivalent of the more familiar invariance under
local transformations that the configuration space action~(\ref{action})
possesses.

\subsection{Gauge-fixed photon}
\label{subsec: Gauge-fixed photon}

One particularly convenient way to fix the gauge is the multiplier gauge,
where we by hand fix the Lagrange multiplier in
the canonical action~(\ref{invariant canonical action}) to be a function of 
canonical variables, which here we choose as,
\begin{equation}
\ell \rightarrow \overline{\ell} = - \frac{\xi a^{4-D}}{2}  \Pi_0 + \partial_i A_i - (D\!-\!2) \mathcal{H} A_0 \, ,
\label{multiplier gauge}
\end{equation}
with~$\xi$ an arbitrary real parameter.
This leads to the gauge-fixed action,
\begin{equation}
\mathscr{S}_\star\bigl[ A_0, \Pi_0, A_i, \Pi_i \bigr]
	\equiv
	\mathscr{S} \bigl[ A_0, \Pi_0, A_i, \Pi_i, \overline{\ell} \, \bigr]
	=
	\int\! d^{D\!}x \, \Bigl[
		\Pi_0 \partial_0 A_0 + \Pi_i \partial_0 A_i - \mathscr{H}_\star
		\Bigr]
		\, ,
\label{gauge fixed action}
\end{equation}
where the gauge-fixed Hamiltonian is,
\begin{equation}
\mathscr{H}_\star = 
	\frac{ a^{4-D} }{2} \Pi_i \Pi_i
	- \frac{a^{4-D} \xi}{2}  \Pi_0 \Pi_0
	- A_0 \partial_i \Pi_i + \Pi_0 \partial_i A_i
	- (D\!-\!2) \mathcal{H} \Pi_0 A_0
	+ \frac{a^{D-4}}{4} F_{ij} F_{ij} \, .
	\ \
\label{gauge fixed Hamiltonian}
\end{equation}
This gauge fixed canonical action now uniquely defines the gauge fixed dynamics,
but it no longer encodes the first-class constraints~(\ref{primary constraint}) and~(\ref{secondary constraint}). 
We have to require them as {\it subsidiary conditions,}
\begin{equation}
\Psi_1 = \Pi_0 \approx 0 \, ,
\qquad \quad
\Psi_2 = \partial_i \Pi_i \approx 0 \, ,
\label{two constraints}
\end{equation}
{\it in addition} to the gauge fixed action. These are preserved if they are demanded on 
the initial time hypersurface.~\footnote{
 This property is guaranteed by the equations of motion the constraints satisfy,
\begin{equation*}
\partial_0\Psi_1 = \Psi_2 + (D\!-\!2)\mathcal{H}\Psi_1 \, ,
\qquad \qquad
\partial_0\Psi_2 = \nabla^2\Psi_1 \, . 
\end{equation*}
}
Thus we split the description of the system into the dynamics described by the gauge-fixed 
action~(\ref{gauge fixed action}), and kinematics given by the subsidiary conditions~(\ref{two constraints}).
This structure will be crucial when quantizing the system.

The utility of the particular choice~(\ref{multiplier gauge}) for the multiplier is 
revealed once we derive the gauge fixed Lagrangian action, that takes the 
form,
\begin{equation}
 S_\star[A_\mu] \!=\! S[A_\mu] + S_{\rm gf}[A_\mu]
 \,,
\label{gauge fixed action}
\end{equation}
where the gauge-fixing term,
\begin{equation}
S_{\rm gf}[A_\mu] = \int\! d^{D\!}x \, \sqrt{-g} \, \biggl[
	- \frac{1}{2\xi} \bigl( g^{\mu\nu} \nabla_\mu A_\nu \bigr)^{\!2} \,
	\biggr] \, ,
\end{equation}
is precisely the general covariant gauge-fixing term. The two subsidiary conditions
substituting for first-class constraints in the Lagrangian formalism take the form,
\begin{equation}
\nabla^\mu A_\mu \approx 0 \, ,
\qquad \qquad
\partial_i F_{0i} \approx 0 \, .
\label{first class constraints: Lagr}
\end{equation}
Implementing these at the initial value surface guarantees they are conserved for all times.

\subsection{Quantized photon}
\label{subsec: Quantized photon}

The gauge fixed photon of the preceding section is readily quantized in the Heisenberg picture.
The dynamics is quantized by applying standard rules of canonical quantization
to the gauge fixed canonical action. 
Canonical variables are promoted to field operators,
and their Poisson brackets~(\ref{Poisson brackets}) are promoted to commutators,
\begin{equation}
\hspace{-2mm}
\bigl[ \hat{A}_0(\eta,\vec{x}) , \hat{\Pi}_0(\eta,\vec{x}^{\,\prime}) \bigr]
	= i \delta^{D-1}(\vec{x} \!-\! \vec{x}^{\,\prime}) ,
\quad
\bigl[ \hat{A}_i(\eta,\vec{x}) , \hat{\Pi}_j(\eta,\vec{x}^{\,\prime}) \bigr]
	= i \delta_{ij} \delta^{D-1}(\vec{x} \!-\! \vec{x}^{\,\prime}) 
	 ,
\label{canonical commutators}
\end{equation}
where we set $\hbar=1$,
and the canonical operator equations of motion,
\begin{align}
\partial_0 \hat{A}_0 ={}&
	- \xi a^{4-D} \hat{\Pi}_0
	+ \partial_i \hat{A}_i
	- (D \!-\! 2 ) \mathcal{H} \hat{A}_0
	\, ,
\label{position eom 1}
\\
\partial_0 \hat{\Pi}_0 ={}&
	\partial_i \hat{\Pi}_i
	+ (D\!-\!2) \mathcal{H} \hat{\Pi}_0
	\, ,
\label{position eom 2}
\\
\partial_0 \hat{A}_i ={}&
	a^{4-D} \hat{\Pi}_i
	+ \partial_i \hat{A}_0
	\, ,
\label{position eom 3}
\\
\partial_0 \hat{\Pi}_i ={}&
	\partial_i \hat{\Pi}_0
	+ a^{D-4} \partial_j \hat{F}_{ji}
	\, ,
\label{position eom 4}
\end{align}
take the same form as in the classical theory.

The constraints~(\ref{two constraints}) of the classical theory require a
more careful quantization.
It is straightforward to associate Hermitian operators associated to classical first-class constraints,
\begin{equation}
\hat{\Psi}_1 = \hat{\Pi}_0 \, ,
\qquad \quad
\hat{\Psi}_2 = \partial_i \hat{\Pi}_i \, ,
\label{Hermitian constraint operators}
\end{equation}
but the constraints clearly cannot be implemented at the operator level, as they would contradict 
canonical commutation relations~(\ref{canonical commutators}). This is of no consequence, since 
the actual property that needs to be satisfied in the quantized theory is that all correlators of the 
Hermitian constraint operators~(\ref{Hermitian constraint operators}) have to vanish. In particular,
for Gaussian states we consider in this paper this means that all the two-point functions
of Hermitian constraints must vanish,
\begin{equation}
\bigl\langle \Omega \bigr|\hat{\Pi}_0(x) \hat{\Pi}_0(x') \bigl| \Omega \bigr\rangle = 0 \, ,
\quad
\bigl\langle \Omega \bigr| \partial_i \hat{\Pi}_i(x) \hat{\Pi}_0(x') \bigl| \Omega \bigr\rangle = 0 \, ,
\quad
\bigl\langle \Omega \bigr| \partial_i \hat{\Pi}_i(x) \partial'_j\hat{\Pi}_j(x') \bigl| \Omega \bigr\rangle = 0 \, .
\label{constraint two-point functions}
\end{equation}
This implies that it is appropriate to implement the constraints as conditions on states. 
The form that such quantum {\it subsidiary conditions} take is not
immediately obvious, since requiring Hermitian constraint operators~(\ref{Hermitian constraint operators}) 
to annihilate the physical state 
would again contradict canonical commutation relations~(\ref{canonical commutators}).
The consistent way of implementing the constraints as subsidiary conditions on states is to
require a non-Hermitian {\it subsidiary constraint operator}~$\hat{K}(\vec{x})$ to annihilate the 
{\it ket} state vector, and its conjugate to annihilate the {\it bra} state vector,
\begin{equation}
\hat{K}(\vec{x}) \bigl| \Omega \bigr\rangle = 0
	\, ,
\qquad \qquad
\bigl\langle \Omega \bigr| \hat{K}^\dag(\vec{x}) = 0 
	\, .
\label{position subsidiary}
\end{equation}
This non-Hermitian constraint operator is constructed as an invertible linear combination of Hermitian constraints,
\begin{equation}
\hat{K}(\vec{x}) = 
	\int\! d^{D-1}x' \, \biggl[
		f_1( \eta,\vec{x} \!-\! \vec{x}^{\,\prime} ) \hat{\Psi}_1(\eta,\vec{x}^{\,\prime})
		+ f_2(\eta,\vec{x} \!-\!  \vec{x}^{\,\prime}) \hat{\Psi}_2(\eta,\vec{x}^{\,\prime})
		\biggr] \, .
\label{position K}
\end{equation}
This guarantees that physical conditions~(\ref{constraint two-point functions}) are always satisfied
in a way consistent with canonical commutation relations~(\ref{canonical commutators}).
The structure of the subsidiary constraint operator becomes clearer when considered in 
momentum space in Sec.~\ref{subsec: Subsidiary condition}.

Since definition~(\ref{position K}) is invertible, it implies that we can decompose the two
Hermitian constraints in terms of the non-Hermitian one and its conjugate. This will be used
in Sec.~\ref{sec: Choice of state} to simplify expressions. It suffices here to define this inverse symbolically as,
\begin{equation}
\hat{\Psi}_1(x) = \hat{K}_1(x) + \hat{K}_1^\dag(x) \, ,
\qquad \qquad
\hat{\Psi}_2(x) = \hat{K}_2(x) + \hat{K}_2^\dag(x) \, ,
\label{nonhermitian split}
\end{equation}
where~$\hat{K}_1(x)$ and~$\hat{K}_2(x)$ are linear in~$\hat{K}(\vec{x})$, and their conjugates are
linear in~$\hat{K}^\dag(\vec{x})$.

\section{Field operator dynamics}
\label{sec: Field operator dynamics}

The free theory defines the propagators for perturbative
computations in the interacting theory.
The dynamics of the free theory is determined by solving
the equations of motion of the field operators, which we do in this section.
The most convenient way to solve for field operators is in momentum space.
After decomposing field operators into transverse and scalar sectors,
we solve for the mode functions in momentum space, and determine the
accompanying commutation relations. The mode functions are expressed in
terms of CTBD scalar mode functions~(\ref{CTBD solution}) and their parametric derivatives.
Our solutions are consistent with previously obtained solutions for photon mode functions
in~$D\!=\!4$~\cite{BeltranJimenez:2008enx,Zhang:2022csl}, and for massive Stueckelberg
vector field mode functions~\cite{Frob:2013qsa}.
The section concludes with a discussion of
momentum space non-Hermitian constraint operators defining the 
subsidiary condition on the space of states.

\subsection{Decomposition of field operators}
\label{subsec: Decomposition of field operators}

It is convenient to split the spatial components of field operators,
\begin{equation}
\hat{A}_i = \hat{A}_i^{\scr T} + \hat{A}_i^{\scr L} \, ,
\qquad \qquad
\hat{\Pi}_i = \hat{\Pi}_i^{\scr T} + \hat{\Pi}_i^{\scr L} \, ,
\label{T L split}
\end{equation}
into transverse parts,~$\hat{A}_i^{\scr T} \!=\! \mathbb{P}_{ij}^{\scr T} \hat{A}_j$ 
and~$\hat{\Pi}_i^{\scr T} \!=\! \mathbb{P}_{ij}^{\scr T} \hat{\Pi}_j$, and longitudinal 
parts,~$\hat{A}_i^{\scr L} \!=\! \mathbb{P}_{ij}^{\scr L} \hat{A}_j$ 
and~$\hat{\Pi}_i^{\scr L} \!=\! \mathbb{P}_{ij}^{\scr L} \hat{\Pi}_j$, 
defined via the projection operators,
\begin{equation}
\mathbb{P}_{ij}^{\scr T} = \delta_{ij} - \frac{\partial_i \partial_j}{\nabla^2} \, ,
\qquad \qquad
\mathbb{P}_{ij}^{\scr L} = \frac{\partial_i \partial_j}{\nabla^2} \, ,
\end{equation}
where~$\nabla^2 \!=\! \partial_i \partial_i$ is the Laplace operator.
The projectors are both idempotent,~$\mathbb{P}_{ij}^{\scr T} \mathbb{P}_{jk}^{\scr T} \!=\! \mathbb{P}_{ik}^{\scr T} $
and~$\mathbb{P}_{ij}^{\scr L} \mathbb{P}_{jk}^{\scr L} \!=\! \mathbb{P}_{ik}^{\scr L} $,
and mutually orthogonal,~$\mathbb{P}_{ij}^{\scr T} \mathbb{P}_{jk}^{\scr L} \!=\! \mathbb{P}_{ij}^{\scr L} \mathbb{P}_{jk}^{\scr T} \!=\! 0 $.
Thus~$\partial_i A_i^{\scr T} \!=\! \partial_i \Pi_i^{\scr T} \!=\! 0$,
and~$\partial_i A_i \!=\! \partial_i A_i^{\scr L}$,~$\partial_i \Pi_i \!=\! \partial_i \Pi_i^{\scr L}$.
Given the spatial isotropy of spatially flat cosmological spaces, it is most convenient to 
examine operator dynamics in the comoving momentum space,
\begin{subequations}
\begin{align}
\hat{A}_0(\eta,\vec{x})
	={}&
	a^{ \frac{2-D}{2} } \!
	\int\! \frac{ d^{D-1}k }{ (2\pi)^{\frac{D-1}{2}} } \,
	e^{i \vec{k} \cdot \vec{x} }
	\, \hat{\mathcal{A}}_0(\eta,\vec{k}) \, ,
\label{mode decompos: A 0}
\\
\hat{\Pi}_0(\eta,\vec{x})
	={}&
		a^{\frac{D-2}{2} } \!
	\int\! \frac{ d^{D-1}k }{ (2\pi)^{\frac{D-1}{2}} } \,
	e^{i \vec{k} \cdot \vec{x} }
	\, \hat{\pi}_{0}(\eta,\vec{k}) \, ,
\\
\hat{A}_i^{\scr L}(\eta,\vec{x})
	={}&
	a^{ \frac{2-D}{2} } \!
	\int\! \frac{ d^{D-1}k }{ (2\pi)^{\frac{D-1}{2}} } \,
	e^{i \vec{k} \cdot \vec{x} }
	\frac{(-i ) k_i}{k}
	\, \hat{\mathcal{A}}_{\scr L}(\eta,\vec{k}) \, ,
\label{mode decompos: A L}
\\
\hat{\Pi}_i^{\scr L}(\eta,\vec{x})
	={}&
	a^{\frac{D-2}{2} } \!
	\int\! \frac{ d^{D-1}k }{ (2\pi)^{\frac{D-1}{2}} } \,
	e^{i \vec{k} \cdot \vec{x} }
	\frac{(-i ) k_i}{k}
	\, \hat{\pi}_{\scr L}(\eta,\vec{k}) \, ,
\\
\hat{A}_i^{\scr T}(\eta,\vec{x})
	={}&
	a^{ \frac{4-D}{2} } \!
	\int\! \frac{ d^{D-1}k }{ (2\pi)^{\frac{D-1}{2}} } \,
	e^{i \vec{k} \cdot \vec{x} }
	\sum_{\sigma=1}^{D-2} \varepsilon_i(\sigma,\vec{k})
	\, \hat{\mathcal{A}}_{{\scr T},\sigma}(\eta,\vec{k}) \, ,
\label{mode decompos: A T}
\\
\hat{\Pi}_i^{\scr T}(\eta,\vec{x})
	={}&
	a^{\frac{D-4}{2} } \!
	\int\! \frac{ d^{D-1}k }{ (2\pi)^{\frac{D-1}{2}} } \,
	e^{i \vec{k} \cdot \vec{x} }
	\sum_{\sigma=1}^{D-2} \varepsilon_i (\sigma,\vec{k})
	\, \hat{\pi}_{{\scr T},\sigma}(\eta,\vec{k}) \, ,
\end{align}
\label{Fourier transforms}%
\end{subequations}
where a Hermitian operator behaves under conjugation
as~$\hat{\mathcal{O}}^\dag(\vec{k}) \!=\! \hat{\mathcal{O}}(-\vec{k})$.
Here we have introduced transverse polarization tensors with the following properties,
\begin{subequations}
\begin{align}
&
\qquad
k_i \, \varepsilon_i(\sigma, \vec{k}) = 0 \, ,
&
\bigl[ \varepsilon_i (\sigma,\vec{k}) \bigr]^*
	= \varepsilon_i(\sigma, - \vec{k}) \, ,
	\qquad
\\
&
\varepsilon_i^* (\sigma,\vec{k}) \, \varepsilon_i(\sigma', \vec{k})
	= \delta_{\sigma \sigma'} \, ,
	\qquad
&
\sum_{\sigma=1}^{D-2} \varepsilon_i^*(\sigma, \vec{k}) \varepsilon_j(\sigma, \vec{k})
	= \delta_{ij} - \frac{ k_i k_j }{ k^2 }
	\, ,
\end{align}
\end{subequations}
where~$k\!=\!\| \vec{k} \|$.
The nonvanishing canonical commutators of the momentum space field operators are,
\begin{subequations}
\begin{align}
&
\bigl[ \hat{\mathcal{A}}_0(\eta,\vec{k} ) , \hat{\pi}_0(\eta,\vec{k}^{\,\prime} ) \bigr]
	= \bigl[ \hat{\mathcal{A}}_{\scr L}(\eta,\vec{k} ) , \hat{\pi}_{\scr L}(\eta,\vec{k}^{\,\prime} ) \bigr]
	= i \delta^{D-1} ( \vec{k} \!+\! \vec{k}^{\,\prime} ) \, ,
\\
&
	\bigl[ \hat{\mathcal{A}}_{{\scr T},\sigma}(\eta,\vec{k} ) , \hat{\pi}_{{\scr T},\sigma'}(\eta,\vec{k}^{\,\prime} ) \bigr]
	= i \delta_{\sigma\sigma'}\delta^{D-1} ( \vec{k} \!+\! \vec{k}^{\,\prime} ) 
	\, .
\end{align}
\label{momentum commutators}%
\end{subequations}
The equations of motion for the transverse sector read,
\begin{align}
\partial_0 \hat{\mathcal{A}}_{{\scr T},\sigma}
	={}&
	\hat{\pi}_{{\scr T},\sigma} + \frac{1}{2} (D\!-\!4) \mathcal{H} \hat{\mathcal{A}}_{{\scr T},\sigma} \, ,
\label{AT eq}
\\
\partial_0 \hat{\pi}_{{\scr T},\sigma}
	={}&
	- k^2 \hat{\mathcal{A}}_{{\scr T},\sigma}
	- \frac{1}{2} (D\!-\!4) \mathcal{H} \hat{\pi}_{{\scr T},\sigma} \, ,
\label{PiT eq}
\end{align}
while the ones for the scalar sector are,
\begin{align}
\partial_0 \hat{\mathcal{A}}_{0}
	={}&
	- \xi a^2 \hat{\pi}_{0}
	+ k \hat{\mathcal{A}}_{\scr L}
	- \frac{1}{2} (D\!-\!2) \mathcal{H} \hat{\mathcal{A}}_{0} 
	\, ,
\label{A0 eq}
\\
\partial_0 \hat{\pi}_{0}
	={}&
	k \hat{\pi}_{\scr L}
	+ \frac{1}{2} (D\!-\!2) \mathcal{H} \hat{\pi}_{0} 
	\, ,
\label{Pi0 eq}
\\
\partial_0 \hat{\mathcal{A}}_{\scr L}
	={}&
	a^2 \hat{\pi}_{\scr L}
	- k \hat{\mathcal{A}}_{0}
	+ \frac{1}{2} (D\!-\!2) \mathcal{H} \hat{\mathcal{A}}_{\scr L}
	\, ,
\label{AL eq}
\\
\partial_0 \hat{\pi}_{\scr L}
	={}&
	- k \hat{\pi}_{0}
	- \frac{1}{2} (D\!-\!2) \mathcal{H} \hat{\pi}_{\scr L}
	\, .
\label{PiL eq}
\end{align}
%

\subsection{Transverse sector}
\label{subsec: Transverse sector}

Equations~(\ref{AT eq}) and~(\ref{PiT eq}) combine into a single second order one,
\begin{align}
\biggl[ \partial_0^2 + k^2 
	- \Bigl( \nu^2 \!-\! \frac{1}{4} \Bigr) \mathcal{H}^2 \biggr]
		\hat{ \mathcal{A} }_{ {\scr T} , \sigma} ={}& 0 
	\, ,
\\
\hat{\pi}_{ {\scr T} ,\sigma} ={}& \biggl[
	\partial_0 - \Bigl( \nu \!+\! \frac{1}{2} \Bigr) \mathcal{H}
	\biggr] \hat{\mathcal{A}}_{ {\scr T} ,\sigma}
	\, ,
\end{align}
where we introduce and henceforth make frequent use of the parameter,
\begin{equation}
\nu = \frac{D\!-\!3 }{ 2 } \, .
\label{nu def}
\end{equation}
The equations are solved by,
\begin{align}
\hat{\mathcal{A}}_{ {\scr T} ,\sigma}(\eta,k)
	={}&
	U_{\nu}(\eta,k) \, \hat{b}_{\scr T}(\sigma, \vec{k})
	+
	U_{\nu}^*(\eta,k) \, \hat{b}_{\scr T}^\dag(\sigma,-\vec{k})
	\, ,
\label{AT final solution}
\\
\hat{\pi}_{ {\scr T} ,\sigma}(\eta,k)
	={}&
	- i k U_{\nu-1}(\eta,k) \, \hat{b}_{\scr T}(\sigma, \vec{k})
	+
	i k U_{\nu-1}^*(\eta,k) \, \hat{b}_{\scr T}^\dag(\sigma,-\vec{k})
	\, .
\end{align}
From the canonical commutation relations~(\ref{momentum commutators}) it then follows that
the initial condition operators
are the creation and annihilation operators
$\hat{b}^\dag_{\scr T} (\sigma, \vec{k})$ and $\hat{b}_{\scr T}(\sigma \! , \vec{k})$, respectively,
whose non-vanishing commutators are,
\begin{equation}
\bigl[ \hat{b}_{\scr T} (\sigma, \vec{k}) , \hat{b}_{\scr T}^\dag(\sigma' \! , \vec{k}^{\,\prime}) \bigr]
	=
	\delta_{\sigma \sigma'} \, \delta^{D-1}(\vec{k} \!-\! \vec{k}^{\,\prime} )
	\, .
\end{equation}
%

\subsection{Scalar sector}
\label{subsec: Scalar sector}

The two equations~(\ref{Pi0 eq}) and~(\ref{PiL eq}) of the scalar sector
decouple, and combine into a single second order one,
\begin{align}
\biggl[ \partial_0^2 + k^2 
	- \Bigl( \nu^2 \!-\! \frac{1}{4} \Bigr) \mathcal{H}^2 \biggr]
		\hat{\pi}_{\scr L} ={}& 0 
	\, ,
\\
\hat{\pi}_0 ={}& - \frac{1}{k} \biggl[
	\partial_0 + \Bigl( \nu\!+\! \frac{1}{2} \Bigr) \mathcal{H}
	\biggr] \hat{\pi}_{\scr L}
	\, ,
\end{align}
where~$\nu$ is given in~(\ref{nu def}),
so that solutions are readily read 
off from~(\ref{mode eq}--\ref{CTBD solution})
and~(\ref{mode recurrence}),
\begin{align}
\hat{\pi}_{\scr L}(\eta, \vec{k}) 
	={}&
	k U_\nu (\eta,k) \hat{b}_{\scr P}(\vec{k})
		+ k U_\nu^*(\eta,k) \hat{b}_{\scr P}^\dag(-\vec{k}) 
	\, ,
\label{piL solution}
\\
\hat{\pi}_{0}(\eta, \vec{k}) 
	={}&
	i k U_{\nu+1} (\eta,k) \hat{b}_{\scr P}(\vec{k})
		- i k U_{\nu+1}^*(\eta,k) \hat{b}_{\scr P}^\dag(-\vec{k}) 
	\, .
\label{pi0 solution}
\end{align}
The remaining equations~(\ref{A0 eq}) and~(\ref{AL eq}) now combine into a second order one
sourced by homogeneous solutions~(\ref{piL solution}) and~(\ref{pi0 solution}),
\begin{align}
\biggl[ \partial_0^2 + k^2 
	- \Bigl( \nu^2 \!-\! \frac{1}{4} \Bigr) \mathcal{H}^2 \biggr]
	\hat{\mathcal{A}}_{0} 
	={}& 
	(1\!-\!\xi) \frac{\mathcal{H}^2 k}{ H^2} \hat{\pi}_{\scr L}
	- 2 \xi \frac{\mathcal{H}^3}{ H^2} \hat{\pi}_0
	\, ,
\\
\hat{\mathcal{A}}_{\scr L} 
	={}&
	\frac{1}{k} \biggl[
		\partial_0 + \Bigl( \nu\!+\! \frac{1}{2} \Bigr) \mathcal{H}
		\biggr] \hat{\mathcal{A}}_{0} 
		+ \frac{ \xi \mathcal{H}^2 }{ H^2 k} \hat{\pi}_0 \, .
\end{align}
The homogeneous parts of the solution are the same as before,
\begin{align}
\hat{\mathcal{A}}_0 (\eta,\vec{k})
	={}&
	U_\nu(\eta,k) \hat{b}_{\scr H}(\vec{k})
	+ U_\nu^*(\eta,k) \hat{b}_{\scr H}^\dag(-\vec{k})
\nonumber \\
&	\hspace{4cm}
	+ v_0(\eta,k) \hat{b}_{\scr P}(\vec{k})
	+ v_0^*(\eta,k) \hat{b}_{\scr P}^\dag(-\vec{k})
	\, ,
\label{A 0 final solution}
\\
\hat{\mathcal{A}}_{\scr L} (\eta,\vec{k})
	={}&
	- i U_{\nu+1}(\eta,k) \hat{b}_{\scr H}(\vec{k})
	+ i U_{\nu+1}^*(\eta,k) \hat{b}_{\scr H}^\dag(-\vec{k})
\nonumber \\
&	\hspace{4cm}
	- i v_{\scr L}(\eta,k) \hat{b}_{\scr P}(\vec{k})
	+ i v_{\scr L}^*(\eta,k) \hat{b}_{\scr P}^\dag(-\vec{k})
	\, ,
\label{A L final solution}
\end{align}
while the particular particular mode functions $v_0$ and $v_{\scr L}$
satisfy,
\begin{align}
\biggl[ \partial_0^2 + k^2 
	- \Bigl( \nu^2 \!-\! \frac{1}{4} \Bigr) \mathcal{H}^2 \biggr]
	v_{0} 
	={}&\! 
	\frac{ - i \xi \mathcal{H}^3 k }{ ( \nu \!+\! 1 ) H^2 }
		\biggl[ 2 ( \nu \!+\! 1 ) U_{\nu+1} 
			- \frac{i k}{\mathcal{H}} U_\nu \biggr]
	+ \Bigl( 1 \!-\! \frac{ \xi }{ \xi_s } \Bigr) \frac{ \mathcal{H}^2 k^2}{ H^2 } U_\nu
	\, ,
\label{particular eq 1}
\\	
v_{\scr L} 
	={}&
	\frac{i}{k} \biggl[
		\partial_0 + \Bigl( \nu\!+\! \frac{1}{2} \Bigr) \mathcal{H}
		\biggr] v_{0} 
		- \frac{\xi \mathcal{H}^2 }{ H^2 } U_{\nu+1} 
		\, ,
\label{particular eq 2}
\end{align}
where we introduced what we refer to as the~{\it simple covariant gauge},
\begin{equation}
\xi_s = \frac{ \nu \!+\! 1 }{ \nu }
	= \frac{ D \!-\! 1 }{ D \!-\! 3 } \, ,
\label{simple covariant gauge}
\end{equation}
which in the flat space limit corresponds to the~$D$-dimensional Fried-Yennie 
gauge~\cite{Fried:1958zz,Adkins:1993qm}.
The photon two-point function takes
 the simplest form in this gauge, as already noted in~\cite{Youssef:2010dw}.
The solutions for the particular mode functions are readily found from identities~(\ref{mode id1})
and~(\ref{mode id2}),
\begin{align}
v_0 ={}&
	\frac{ - i \xi k}{ 2 (\nu\!+\!1) H } 
		\biggl[ \frac{ \mathcal{H} }{ H } U_{\nu+1} - U_{\nu} \biggr]
	- 
	\Bigl( 1 \!-\! \frac{ \xi }{ \xi_s } \Bigr)
	\frac{ i k }{ 2 H }
	\biggl[
	\frac{ i k }{ \nu H} \frac{\partial U_{\nu} }{ \partial \nu } + U_{\nu}
	\biggr]
	\, ,
\label{dS particular mode 0}
\\
v_{\scr L} ={}&
	\frac{ - i \xi k}{ 2 (\nu\!+\!1) H } 
		\biggl[ 
		\frac{ \mathcal{H} }{ H } U_{\nu} 
		\! - U_{\nu+1} 
		\biggr]
	- 
	\Bigl( 1 \!-\! \frac{ \xi }{ \xi_s } \Bigr)
	\frac{ i k }{ 2 H }
	\biggl[
		\frac{ i k }{ \nu H}  \frac{\partial U_{\nu+1} }{ \partial \nu } 
		\! + \frac{ \mathcal{H} }{ \nu H} U_{\nu} \!
		+ U_{\nu+1}
	\biggr]
	\, ,
\label{dS particular mode L}
\end{align}
where the homogeneous parts were fixed by requiring the Wronskian-like relation,
\begin{equation}
{\rm Re} \Bigl( v_0 U_{\nu+1}^* + v_{\scr L} U_\nu^* \Bigr) = 0 \, ,
\label{Wronskian-like}
\end{equation}
and a regular flat space limit,
\begin{subequations}
\begin{align}
v_0 \xrightarrow{H\to0}{}&
	\frac{1}{4}  \Bigl[ (1\!+\!\xi) \!+\! 2(1\!-\!\xi)ik (\eta\!-\!\eta_0) \Bigr] 
	\frac{ e^{-ik(\eta-\eta_0)} }{ \sqrt{2k} } \, ,
\\
v_{\scr L} \xrightarrow{H\to0}{}&
	\frac{1}{4} \Bigl[ - (1\!+\!\xi) + 2 (1\!-\!\xi) i k(\eta \!-\! \eta_0)  \Bigr]
	\frac{ e^{-ik(\eta-\eta_0)} }{ \sqrt{2k} } \, .
\end{align}
\label{mode flat limit}%
\end{subequations}
Using the Wronskian-like relation above we can invert the momentum space field operators
for the initial conditions operators, and infer their commutation relations
from the canonical ones~(\ref{momentum commutators}),
with the non-vanishing ones being,
\begin{equation}
\bigl[ \hat{b}_{\scr H}(\vec{k}) , \hat{b}_{\scr P}^\dag(\vec{k}^{\,\prime}) \bigr]
	=
	\bigl[ \hat{b}_{\scr P}(\vec{k}) , \hat{b}_{\scr H}^\dag(\vec{k}^{\,\prime}) \bigr]
	= 
	- \delta^{D-1}( \vec{k} \! - \! \vec{k}^{\,\prime})
	 \, .
\label{bH bP commutator}
\end{equation}
These commutators are not canonical, in the sense that they are not the ones of creation/annihilation
operators. It is actually advantageous to keep them in this form for the purpose of
constructing the space of states, as it simplifies matters and makes the structure
more transparent~\cite{Glavan:2022a}.

\subsection{Subsidiary condition}
\label{subsec: Subsidiary condition}

Physical states have to to satisfy the subsidiary condition~(\ref{position subsidiary}) descending 
from the first-class constraints. For this purpose we need to construct the non-Hermitian constraint 
operator~(\ref{position K}). It takes a considerably simpler, diagonal
form in momentum space, 
\begin{equation}
\hat{K}(\vec{x}) = a^{\frac{D-2}{2}} \int\! \frac{d^{D-1}k}{(2\pi)^{\frac{D-1}{2}} } \, e^{i\vec{k}\cdot\vec{x}} \, \hat{\mathcal{K}}(\vec{k}) \, ,
\qquad
\hat{K}^\dag(\vec{x}) = a^{\frac{D-2}{2}} \int\! \frac{d^{D-1}k}{(2\pi)^{\frac{D-1}{2}} } \, e^{i\vec{k}\cdot\vec{x}} \, \hat{\mathcal{K}}^\dag(-\vec{k}) \, ,
\end{equation}
where it
is a simple linear combination of the momentum space Hermitian constraints,
\begin{equation}
\hat{\mathcal{K}}(\vec{k})
	= c_0(\eta,\vec k) \hat{\pi}_0(\eta,\vec{k})
	+ c_{\scr L}(\eta,\vec k) \hat{\pi}_{\scr L}(\eta,\vec{k})
	\, ,
\label{momentum K}
\end{equation}
so that the equivalent subsidiary conditions are,
\begin{equation}
\hat{\mathcal{K}}(\vec{k}) \bigl| \Omega \bigr\rangle = 0 \, ,
\qquad \quad
\bigl\langle \Omega \bigr| \hat{\mathcal{K}}^\dag(\vec{k}) = 0 
\, ,
\qquad \quad
\forall \vec{k} \, .
\label{momentum subsidiary}
\end{equation}
The non-Hermitian constraint is time-independent, which implies the following 
first order equations of motion for the coefficient functions,
\begin{equation}
\partial_0 c_0 - k c_{\scr L} + \frac{1}{2} (D\!-\!2) \mathcal{H} c_0 = 0 \, ,
\qquad \quad
\partial_0 c_{\scr L} + k c_0 - \frac{1}{2} (D\!-\!2) \mathcal{H} c_{\scr L} = 0 \, .
\end{equation}
These equations combine into a second order one,
\begin{align}
\biggl[ \partial_0^2 + k^2 
	- \Bigl( \nu^2 \!-\! \frac{1}{4} \Bigr) \biggr] c_0 ={}& 0 \, ,
\\
c_{\scr L} ={}&
	\frac{1}{k} \biggl[ \partial_0 
		+ \Bigl( \nu\!+\! \frac{1}{2} \Bigr) \mathcal{H} \biggr] c_0 \, ,
\end{align}
whose general solution, according to~(\ref{mode eq}--\ref{CTBD solution}),
can be written as,
\begin{align}
c_0(\eta,\vec{k}) ={}& \alpha(\vec{k}) U_\nu(\eta,k) + \beta(\vec{k}) U_\nu^*(\eta,k) \, ,
\\
c_{\scr L}(\eta,\vec{k}) ={}& - i \alpha(\vec{k}) U_{\nu+1}(\eta,k) + i \beta(\vec{k}) U_{\nu+1}^*(\eta,k) \, ,
\end{align}
where~$\alpha(\vec{k})$ and~$\beta(\vec{k})$ are some free functions of momenta with immaterial normalization.
We parametrize these coefficient functions so that they are consistent with homogeneity and 
isotropy~\footnote{ A more general 
parametrization is considered in~\cite{Glavan:2022a},
\begin{equation*}
\hat{\mathcal{K}}(\vec{k}) = 
	\mathcal{N}(\vec{k})
	{\rm e}^{i\theta(\vec{k})} \Bigl(
	{\rm e}^{ - i \varphi(\vec{k}) } {\rm ch} [\rho(\vec{k})] \, \hat{b}_{\scr P}(\vec{k})
	+ {\rm e}^{i \varphi(-\vec{k})} {\rm sh}[\rho(-\vec{k})] \, \hat{b}_{\scr P}^\dag(-\vec{k})
	\Bigr)
	\,,
\label{more general parametrization of K}
\end{equation*}
with~$\mathcal{N}(\vec{k}) \!=\! \bigl( {\rm ch}[\rho(\vec{k})] {\rm ch}[\rho(-\vec{k})] 
	\!-\! {\rm sh}[\rho(\vec{k})] {\rm sh}\rho[(-\vec{k})] \bigr)^{-\frac{1}{2}}$, that reduces to the 
one in~(\ref{general ansatz for K}) when homogeneity and isotropy are imposed.
}
 of cosmological
spacetimes~\cite{Glavan:2022a}, such that the non-Hermitian constraint reads,~\footnote{It is also 
possible to exchange the roles of~$\hat{b}_{\scr P}(\vec k)$ and~$\hat{b}_{\scr P}^\dag(\vec k)$. 
The entire analysis can be repeated. However, that choice would not be consistent with
the requirements imposed in Sec.~\ref{sec: Choice of state}.}
\begin{equation}
\hat{\mathcal{K}}(\vec{k}) = {\rm e}^{i\theta(\vec{k})} \Bigl(
	{\rm e}^{ - i \varphi(k) } {\rm ch} [\rho(k)] \, \hat{b}_{\scr P}(\vec{k})
	+ {\rm e}^{i \varphi(k)} {\rm sh}[\rho(k)] \, \hat{b}_{\scr P}^\dag(-\vec{k})
	\Bigr)
	\, ,
\label{general ansatz for K}
\end{equation}
where~$\theta(\vec{k})$,~$\varphi(k)$, and~$\rho(k)$ are real functions of momentum. 
It is also convenient to define an accompanying operator,
\begin{equation}
\hat{\mathcal{B}}(\vec{k})
	= 
	e^{ i \theta(\vec{k})} \Bigl( e^{ - i \varphi(k)} {\rm ch}[ \rho(k) ] \, \hat{b}_{\scr H}(\vec{k}) 
		+ e^{ i \varphi(k)} {\rm sh}[ \rho(k) ] \, \hat{b}_{\scr H}^\dag(-\vec{k}) \Bigr)
		\, ,
\end{equation}
that preserves commutation relations~(\ref{bH bP commutator}),
\begin{equation}
\bigl[ \hat{\mathcal{K}}(\vec{k}) , \hat{\mathcal{B}}^\dag(\vec{k}^{\,\prime}) \bigr]
	=
	\bigl[ \hat{\mathcal{B}}(\vec{k}) , \hat{\mathcal{K}}^\dag(\vec{k}^{\,\prime}) \bigr]
	=
	- \delta^{D-1}(\vec{k} \!-\! \vec{k}^{\,\prime})
	\, ,
\label{algebra}
\end{equation}
so that in this sense these definitions can be seen as Bogolyubov transformations 
between~$\hat{b}_{\scr P}$ and~$\hat{b}_{\scr H}$, and~$\hat{\mathcal{K}}$ and~$\hat{\mathcal{B}}$.
We have to construct the space of states on which 
the algebra in~(\ref{algebra}) is represented, such that it contains the kernel of~$\hat{\mathcal{K}}$ as a subspace. The following section is devoted to this
task.

\section{Constructing the state}
\label{sec: Choice of state}

The quantization of the dynamics outlined in the preceding two sections needs to be supplemented
by the construction of the space of states. Its construction in quantum field theory is necessarily
intertwined with the possible existence of global symmetries that the vacuum 
is sought to respect.
This section presents a detailed analysis of both physical and gauge-fixed de Sitter symmetries 
of the photon in the general covariant gauge. Physical symmetries are related to the gauge-invariant
action~(\ref{action}) and are a property of physical polarizations only, while gauge-fixed symmetries
are related to the gauge-fixed action~(\ref{gauge fixed action}) and are a 
property of the physical and gauge sectors alike.
We define the vacuum state in momentum space mode-by-mode, by requiring the state (i) to
produce vanishing expectation values of all the physical de Sitter generators, and 
(ii) to be an eigenstate of all the gauge-fixed de Sitter generators, with vanishing eigenvalues.
This is a rather involved section and for readers not interested in the technical details of the
procedure we provide a brief summary of the results in the concluding
part~\ref{subsec: Momentum space de Sitter invariant state} of this section.

{
\allowdisplaybreaks

\subsection{De Sitter symmetries}
\label{subsec: De Sitter symmetries}

There are~$\frac{1}{2}D(D\!+\!1)$ isometries of the Poincar\'{e} patch of de Sitter space
(see {\it e.g.}~\cite{Park:2011ww}). 
These are the
coordinate transformations that leave the de Sitter line element invariant.
There is the same number of corresponding active vector field
transformations that are symmetries of both the gauge invariant~(\ref{action}) and the 
gauge fixed action~(\ref{gauge fixed action}). The isometries and the associated
infinitesimal active field transformations are:
\begin{itemize}
\item
{\it Spatial translations} --- $(D\!-\!1)$ transformations,
\begin{equation}
\eta \longrightarrow \eta \, ,
\qquad \quad
x_i \longrightarrow x_i + \alpha_i \, ,
\end{equation}
with the associated active field transformation,
\begin{equation}
A_\mu(x) \longrightarrow A_\mu(x)
	- \alpha_i \partial_i A_\mu(x)
	\, .
\end{equation}

\item
{\it Spatial rotations} --- $\frac{1}{2}(D\!-\!1)(D\!-\!2)$ transformations,
\begin{equation}
\eta \longrightarrow \eta \, ,
\qquad \quad
x_i \longrightarrow x_i + 2 \omega_{ij} x_j \, ,
\qquad \quad
( \omega_{ij} =- \omega_{ji} )
\, ,
\end{equation}
with the associated active field transformation,
\begin{equation}
A_\mu(x) \longrightarrow A_\mu(x)
	+ 2 \omega_{ij} x_i \partial_j A_\mu(x)
	+ 2 \delta_\mu^i \omega_{ij} A_j(x)
	\, .
\end{equation}

\item
{\it Dilation} --- one transformation,
\begin{equation}
\eta \longrightarrow
	\eta - \frac{ \delta }{a}
	\, ,
\qquad \quad
x_i \longrightarrow
	( 1 \!+\! H \delta ) x_i 
	\, ,
\label{dilation}
\end{equation}
with the associated active field transformation,
\begin{equation}
A_\mu(x) \longrightarrow A_\mu(x)
	+ \frac{\delta}{a} \partial_0 A_\mu(x)
	- \delta H x_i \partial_i A_\mu(x)
	- \delta H A_\mu(x)
	\, .
\label{special trans}
\end{equation}

\item
{\it Spatial special conformal transformations} --- $(D\!-\!1)$ transformations,
\begin{equation}
\eta \longrightarrow
	\eta + \frac{ \theta_j x_j  }{a}
	\, ,
\qquad \quad
x_i \longrightarrow
	x_i - H \theta_j x_j x_i
	- \frac{H \theta_i}{2} \biggl[ \frac{1}{H^2} \Bigl( \frac{1}{a^2} \!-\!1 \Bigr) - x_j x_j \biggr]
	\, ,
\end{equation}
with the associated active field transformation,
\begin{equation}
A_\mu(x) \longrightarrow A_\mu(x)
	- \frac{ \theta_i x_i }{a} F_{0\mu}(x)
	+ \biggl\{ H \theta_j x_j x_i
		+ \frac{ H \theta_i }{2} \biggl[ \frac{1}{H^2} \Bigl( \frac{1}{a^2} \!-\! 1 \Bigr) 
			- x_j x_j \biggr] \biggr\} \partial_i A_\mu(x) \, .
\end{equation}

\end{itemize}
Note that the transformations above are written in such a way that in the Minkowski 
limit,~$H\!\to\!0$, they reduce to the Poincar\'{e} transformations. In particular, 
spatial translations and rotations do not change, while
dilation~(\ref{dilation}) reduces to time translation, and spatial special conformal
transformation~(\ref{special trans}) reduces to Lorentz boosts.

In gauge theories one should distinguish between two sets of global symmetry generators --- 
one set descending from the gauge-invariant action~(\ref{gauge invariant action}) 
that accounts for the physical symmetries of
the system, and another set descending from the gauge-fixed action~(\ref{gauge fixed action})
that accounts for the symmetries of
dynamics of the gauge-fixed system. The two sets in general do not have to contain the same number of
generators since gauge-fixing is allowed to break global symmetries of the gauge-invariant system.
In the case at hand both the gauge-invariant action and the gauge-fixed action are invatiant under
all the de Sitter symmetry transformations
above. Correspondingly, the two sets of generators
we compute next are of the same dimensionality.

\subsection{Physical symmetries}
\label{subsec: Physical symmetries}

Physical symmetries of the photon in de Sitter are accounted for by the symmetry generators
descending from the gauge-invariant action~(\ref{action}),
\begin{align}
P_i ={}& 
	\int\! d^{D-1}x \, \Bigl( - \Pi_j F_{ij} \Bigr)
	\, ,
\label{Pi}
\\
M_{ij} ={}&
	\int\! d^{D-1}x \, \Bigl( 2 x_{[i} F_{j]k} \Pi_k \Bigr)
	\, ,
\label{Mij}
\\
Q ={}& \int\! d^{D-1}x \, \biggl(
		\frac{ a^{3-D} }{2a} \Pi_{i} \Pi_{i} 
		+ \frac{ a^{D-5} }{4} F_{ij} F_{ij} 
		- H x_i F_{ij} \Pi_{j} 
		\biggr)
		\, ,
\label{Q}
\\
K_i ={}&
	\int \! d^{D-1}x \, 
	\biggl[
	- \frac{ x_i }{2a} \biggl( a^{4-D} \Pi_{j} \Pi_{j} 
			+ \frac{ a^{D-4}  }{2} F_{jk} F_{jk} \biggr)
		+ H x_i x_j F_{jk} \Pi_{k} 
\nonumber \\
&	\hspace{2.5cm}
	+ \frac{ 1 }{2H} \biggl( \frac{1}{a^2} - 1 
		- H^2 x_k x_k \biggr) F_{ij} \Pi_{j}  
	\biggr]
	\, ,
\label{Ki}
\end{align}
and they satisfy the de Sitter algebra weakly (up to constraints),
\begin{align}
&
\bigl\{ P_i , P_j \bigr\} \approx 0 \, , 
\qquad
\bigl\{ P_i , M_{ij} \bigr\} \approx 2 P_{[i} \delta_{j]k} \, ,
\qquad
\bigl\{ M_{ij} , M_{kl} \bigr\} \approx 4 \delta_{i][k} M_{l][j} \, ,
\nonumber \\
&
\bigl\{ Q , P_i \bigr\} \approx H P_i \, ,
\qquad
\bigl\{ Q , M_{ij} \bigr\} \approx 0 \, ,
\qquad
\bigl\{ Q , Q \bigr\} \approx 0 \, ,
\nonumber \\
&
\bigl\{ K_i , P_j \bigr\} \approx - \delta_{ij} Q - H M_{ij} \, ,
\qquad
\bigl\{ K_i , M_{jk} \bigr\} \approx 2 \delta_{i[j} K_{k]} \, ,
\nonumber \\
&
\bigl\{ K_i , Q \bigr\} \approx - P_i + H K_i \, ,
\qquad
\bigl\{ K_i , K_j \bigr\} \approx M_{ij} \, .
\label{dS algebra}
\end{align}
In the flat space limit,~$H \!\to\! 0$, this algebra reduces to the Poincar\'e algebra.
Quantizing the generators requires one to address operator ordering. Namely,
every term containing a classical constraint should be ordered so that the non-Hermitian
constraint operator is on the right of the product, and that its conjugate is on the left.
This is accomplished by writing them as ({\it cf.} Eqs.~(\ref{Appendix A: Pi gi}),
(\ref{Appendix A: Mij gi}), (\ref{Appendix A: Q gi}) and~(\ref{Appendix A: Ki gi})
from Appendix~\ref{Scalar-transverse decomposition of charges}),
\begin{align}
\hat{P}_i ={}&
	\hat{P}_i^{\scr T}
	+
	\int\! d^{D-1}x \, 
	\Bigl(
	\hat{K}_2^\dag \hat{A}_i^{\scr T}
	+ \hat{A}_i^{\scr T} \hat{K}_2 
	\Bigr)
	\, ,
\label{hatPi}
\\
\hat{M}_{ij} ={}&
	\hat{M}_{ij}^{\scr T}
	+
	\int\! d^{D-1}x \, 
	\Bigl(
	2 \hat{K}_2^\dag x_{[i} \hat{A}_{j]}^{\scr T} 
	+
	2 x_{[i} \hat{A}_{j]}^{\scr T} \hat{K}_2
	\Bigr)
	\, ,
\label{hatMij}
\\
\hat{Q} ={}&
	\hat{Q}^{\scr T}
	+
	\int\! d^{D-1}x
	\biggl(
	- \frac{ a^{3-D} }{2} \hat{\Psi}_2 \nabla^{-2} \hat{\Psi}_2
	- H \hat{K}_2^\dag x_i \hat{A}_i^{\scr T}
	- H x_i \hat{A}_i^{\scr T} \hat{K}_2
	\biggr)
	\, ,
\label{hatQ}
\\
\hat{K}_i ={}&
	\hat{K}_i^{\scr T}
	+
	\int\! d^{D-1}x
	\biggl[
		\frac{a^{D-3}}{2}x_i\hat\Psi_2  \frac{1}{\nabla^2} \hat\Psi_2
		+a^{3-D} \biggl( \hat K_2^\dag\frac{1}{\nabla^2}\Pi_i^{\scr T}
	     		+ \hat\Pi_i^{\scr T}\frac{1}{\nabla^2} \hat K_2 \biggr)
\nonumber \\
&	\hspace{2.cm}	                     
	+(D\!-\!3) H \biggl( \hat{K}_2^\dag \frac{1}{\nabla^2} \hat{A}_i^{\scr T} 
		+ \hat{A}_i^{\scr T} \frac{1}{\nabla^2} \hat{K}_2 \biggr)
	+ H x_i x_j  \Bigl( \hat{K}_2^\dag \hat{A}_j^{\scr T} 
		+ \hat{A}_j^{\scr T} \hat{K}_2 \Bigr)
\nonumber \\
&	\hspace{3cm}
	+ \frac{ 1 }{2H} \biggl( \frac{1}{a^2} \!-\! 1 - H^2 x_j x_j \biggr) 
		\Bigl( \hat{K}_2^\dag \hat{A}_i^{\scr T}
			+ \hat{A}_i^{\scr T} \hat{K}_2 \Bigr)
	\biggr]
	\, ,
\label{hatKi}
\end{align}
where the purely transverse parts are computed to be
({\it cf.} Eqs.~(\ref{Appendix A: Pi T}),
(\ref{Appendix A: Mij T}), (\ref{Appendix A: Q T}) and~(\ref{Appendix A: Ki T})
from Appendix~\ref{Scalar-transverse decomposition of charges}),
\begin{align}
\hat{P}_i^{\scr T} ={}&
		\int\! d^{D-1}k \, k_i \, \hat{\mathcal{E}}_j^\dag(\vec{k}) \hat{\mathcal{E}}_j(\vec{k})
		=
	\int\! d^{D-1}k \, k_i \, \sum_{\sigma=1}^{D-2} \hat{b}_{\scr T}^\dag(\sigma,\vec{k}) \, 
		\hat{b}_{\scr T}(\sigma,\vec{k})
		\, ,
\label{PiT}
\\
\hat{M}_{ij}^{\scr T} ={}&
\int\! d^{D-1} k
	\biggl[
	\hat{\mathcal{E}}_k^\dag(\vec{k})
	\biggl( i k_i \frac{\partial}{\partial k_j} \!-\! i k_j \frac{\partial}{\partial k_i} \biggr)
	\hat{\mathcal{E}}_k(\vec{k})
	+
	2 \hat{\mathcal{E}}_{[i}^\dag(\vec{k}) \hat{\mathcal{E}}_{j]}(\vec{k})
	\biggr]
	\, , 
\label{MijT}
\\
\hat{Q}^{\scr T} ={}&
	\int\! d^{D-1} k \, \biggl[
		k \hat{\mathcal{E}}_i^{\dag}(\vec{k}) \hat{\mathcal{E}}_i(\vec{k})
		- \frac{i H k_j}{2} 
		\biggl(
			\frac{\partial \hat{\mathcal{E}}_i^\dag(\vec{k} )}{ \partial k_j} \hat{\mathcal{E}}_i(\vec{k})
			- 
			\hat{\mathcal{E}}_i^\dag(\vec{k}) \frac{\partial \hat{\mathcal{E}}_i(\vec{k} )}{ \partial k_j}
			\bigg)
		\biggr]
		\, ,
\label{QT}
\\
\hat{K}_i^{\scr T} ={}&
	\int\! d^{D-1}k \, \Biggl[
	\frac{ik}{2} 
		\biggl(
		\frac{\partial \hat{\mathcal{E}}_j^\dag(\vec{k}) }{\partial k_i } \hat{\mathcal{E}}_j(\vec{k})
		-
		\hat{\mathcal{E}}_j^\dag(\vec{k}) \frac{\partial \hat{\mathcal{E}}_j(\vec{k}) }{\partial k_i } 
		\biggr)
	- \frac{ (D\!-\!3)^2 H k_i}{8k^2} 
		\hat{\mathcal{E}}_j^\dag(\vec{k}) \hat{\mathcal{E}}_j(\vec{k})
\nonumber \\
&
	+ \frac{H}{2} 
		\biggl(
		\frac{\partial \hat{\mathcal{E}}_i^\dag(\vec{k}) }{\partial k_j } \hat{\mathcal{E}}_j(\vec{k})
		+
		\hat{\mathcal{E}}_j^\dag(\vec{k}) \frac{\partial \hat{\mathcal{E}}_i(\vec{k}) }{\partial k_j }
		-
		\hat{\mathcal{E}}_i^\dag(\vec{k}) \frac{\partial \hat{\mathcal{E}}_j(\vec{k}) }{\partial k_j }
		-
		\frac{\partial \hat{\mathcal{E}}_j^\dag(\vec{k}) }{\partial k_j } \hat{\mathcal{E}}_i(\vec{k})
		\biggr)
\nonumber \\
&		+ 2 k_i \frac{ \partial \hat{\mathcal{E}}_k^\dag(\vec{k}) }{ \partial k_j } 
			\frac{ \partial \hat{\mathcal{E}}_k(\vec{k}) }{ \partial k_j } 
		-
		2 k_j \frac{ \partial \hat{\mathcal{E}}_k^\dag(\vec{k}) }{ \partial k_j } 
			\frac{ \partial \hat{\mathcal{E}}_k(\vec{k}) }{ \partial k_i } 
		-
		2 k_j \frac{ \partial \hat{\mathcal{E}}_k^\dag(\vec{k}) }{ \partial k_i } 
			\frac{ \partial \hat{\mathcal{E}}_k(\vec{k}) }{ \partial k_j }
		\Biggr]
		\, ,
\label{KiT}
\end{align}
with the expressions written compactly using a short-hand notation,
\begin{equation}
\hat{\mathcal{E}}_i(\vec{k})
	=
	\sum_{\sigma=1}^{D-2} \varepsilon_i(\sigma,\vec{k}) \, \hat{b}_{\scr T}(\sigma,\vec{k}) 
	\, ,
\label{B def}
\end{equation}
and where products have been normal ordered in the usual manner.
Since none of the generators~(\ref{hatPi}--\ref{hatKi}) depends on either~$A_0$ or~$A_i^{\scr L}$,
they automatically qualify as observables since all of them
commute with the non-Hermitian constraint,
\begin{equation}
\bigl[ \hat{\mathcal{K}}(\vec{k}) , \hat{G}_I \bigr] = 0 \, ,
\qquad  \qquad
\hat{G}_I =\{\hat{P}_i, \hat{M}_{ij}, \hat{Q}, \hat{K}_i \}
 \, ,
\end{equation}
for any choice of the parameters in~(\ref{general ansatz for K}).



Firstly, we want to specify a {\it physically} de Sitter-invariant state. 
This is not
implemented as in theories without constraints by considering eigenstates of
the generators, but rather by computing expectation values of polynomials
of generators and showing that the values factorize. In other words, this means,
\begin{equation}
\bigl\langle \Omega \bigr| \mathscr{P} \bigl( \hat{P}_i, \hat{M}_{ij} , \hat{Q}, \hat{K}_i \bigr) 
	\bigl| \Omega \bigr\rangle
	= \mathscr{P} \bigl( \overline{P}_i, \overline{M}_{ij}, \overline{Q}, \overline{K}_i \bigr) \, ,
\label{physical invariance}
\end{equation}
where the arguments on the right-hand-side are expectation values of generators,
\begin{equation}
\overline{P}_i = \bigl\langle \Omega \bigr| \hat{P}_i \bigl| \Omega \bigr\rangle \, ,
\quad
\overline{M}_{ij} =  \bigl\langle \Omega \bigr|  \hat{M}_{ij}  \bigl| \Omega \bigr\rangle
 \, ,
\quad
\overline{Q} = \bigl\langle \Omega \bigr| \hat{Q} \bigl| \Omega \bigr\rangle \, ,
\quad
\overline{K}_i = \bigl\langle \Omega \bigr| \hat{K}_i \bigl| \Omega \bigr\rangle \, .
\label{generator expectations}
\end{equation}
Making this consistent with the generator algebra~(\ref{dS algebra}) then
implies that all of the expectation values of generators~(\ref{generator expectations}) 
have to vanish,
\begin{equation}
\overline{P}_i = 0 \, ,
\qquad
\overline{M}_{ij} = 0 \, ,
\qquad
\overline{Q}= 0 \, , 
\qquad
\overline{K}_i =0 \, .
\label{all zeros}
\end{equation}
Furthermore, only the fully transverse 
parts of the generators in~(\ref{physical invariance}) participate, as any parts
containing scalar sector operators annihilate either the {\it ket} or the {\it bra} state,
\begin{equation}
\bigl\langle \Omega \bigr| \mathscr{P} \bigl( \hat{P}_i, \hat{M}_{ij}, \hat{Q}, \hat{K}_i \bigr) 
	\bigl| \Omega \bigr\rangle
	= \bigl\langle \Omega \bigr| \mathscr{P} \bigl( \hat{P}_i^{\scr T},
		\hat{M}_{ij}^{\scr T}, \hat{Q}^{\scr T} , \hat{K}_i^{\scr T} \bigr) 
	\bigl| \Omega \bigr\rangle
	\, .
\end{equation}
This means that requiring the state to be physically de Sitter symmetric
puts conditions on the transverse sector only.
 Clearly~(\ref{all zeros}) is satisfied if we define the state by,
\begin{equation}
\hat{b}_{\scr T}(\sigma, \vec{k}) \bigl| \Omega \bigr\rangle = 0
 \, ,
 \qquad \quad
 \forall \vec{k}, \sigma \, .
\label{transverse cond on state}
\end{equation}
The rest of the Hilbert space for the transverse sector is then defined in the usual manner as a Fock space,
by acting on~$\bigl| \Omega \bigr\rangle$ with the transverse creation operators. 
It should be noted that de Sitter symmetries do not uniquely fix the state to satisfy~(\ref{transverse cond on state}).
In fact they allow for a two-parameter class of states analogous to~$\alpha$-vacua
of the scalar~\cite{Mottola:1984ar,Allen:1985ux}.
Nevertheless, we consider~(\ref{transverse cond on state}) to define our state, as it  minimizes energy and, in addition, 
it is the unique state with a regular
flat space limit.

It is important to note that the gauge sector of the space of states is not fixed 
by the requirement of the physical de Sitter invariance, and can in fact be chosen freely
without interfering with it. However, it is be advantageous to
 fix the gauge sector to also respect de Sitter symmetries.
 This is discussed and implemented in the following subsection.

\subsection{Gauge-fixed symmetries}
\label{subsec: Manifest de Sitter invariance}

The dynamics of the gauge-fixed theory is determined by the gauge-fixed action~(\ref{gauge fixed action}).
We have chosen the gauge-fixing term to respect general covariance, and hence it respects
de Sitter symmetries of Sec.~\ref{subsec: De Sitter symmetries}. Therefore, the full gauge-fixed action respects de Sitter symmetries,
and accordingly the associated symmetry generators are,
\begin{align}
P_i^\star ={}&
	\int\! d^{D-1}x\,
	\Bigl( - \Pi_j \partial_i A_j
	- \Pi_0 \partial_i A_0 \Bigr)
	\, ,
\label{Pi star}
\\
M_{ij}^\star ={}&
	\int\! d^{D-1}x\, 
	\Bigl(
	2 \Pi_0 x_{[i} \partial_{j]} A_0
	+ 2 \Pi_k x_{[i} \partial_{j]} A_k
	+ 2 \Pi_{[i} A_{j]}
	\Bigr)
	\, ,
\label{Mij star}
\\
Q^\star ={}&
	\int\! d^{D-1}x \, 
	\Biggl\{ \frac{1}{a} \biggl[
		\frac{a^{4-D}}{2} \Pi_i \Pi_i 
		- \frac{a^{4-D} \xi}{2} \Pi_0^2
		- A_0 \partial_i \Pi_i + \Pi_0 \partial_i A_i
		- (D\!-\!2) \mathcal{H} \Pi_0 A_0
\nonumber  \\
&	\hspace{2.cm}
		+ \frac{ a^{D-4} }{ 4 } F_{ij} F_{ij}
		\biggr]
	- H \biggl[ \Pi_j  \bigl( 1 \!+\! x_i \partial_i \bigr) A_j
		+ \Pi_0 \bigl( 1 \!+\! x_i \partial_i \bigr) A_0  \biggr]
		\Biggr\}
		\, ,
\label{Q star}
\\
K_i^\star ={}& \!\!
	\int \!d^{D-1}x \, \Biggl\{
	- \frac{x_i}{a} \biggl[
		\frac{ a^{4-D} }{2} \Pi_j \Pi_j
		- \frac{a^{4-D} \xi}{2} \Pi_0^2
		- A_j \partial_j \Pi_0 
		- A_0 \partial_j\Pi_{j}
		- (D\!-\!2)\mathcal{H} \Pi_0 A_0
\nonumber \\
&	\hspace{-0.5cm}
		+ \frac{a^{D-4}}{4} F_{jk} F_{jk}
		\biggr]
	+ H x_j A_j \Pi_{i}
	+ H \biggl[ x_i x_k \partial_k A_j 
		+ \frac{ 1 }{ 2 H^2 } \biggl( \frac{1}{a^2} \!-\! 1
			\!-\! H^2 x_k x_k \biggr) \partial_i A_j 
\nonumber \\
&	\hspace{-0.5cm}
		+ x_i A_j - x_j A_i
			\biggr] \Pi_{j}
	+ H \biggl[ x_i x_j \partial_j A_0 
		+ \frac{ 1 }{ 2 H^2 } \biggl( \frac{1}{a^2} \!-\! 1
			\!-\! H^2 x_j x_j \biggr) \partial_i A_0 
	+ x_i A_0 
	\biggr] \Pi_0 
		\Biggr\} \, .
\label{Ki star}
\end{align}
These generators satisfy de Sitter algebra~(\ref{dS algebra}) strongly, since the gauge-fixed action
does not possess local symmetries any more. Quantizing these generators requires us to address
operator ordering. The purely transverse part will be the same as for the gauge-invariant generators,
while the remainder warrants a closer look. We need to order operators so that all the non-Hermitian
constraints are on the right in the products, and their conjugates on the left. This is a 
 tedious task, and here we report the results
({\it cf.} Eqs.~(\ref{Appendix A: Pi gf}),
(\ref{Appendix A: Mij gf}), (\ref{Appendix A: Q gf}) and~(\ref{Appendix A: Ki gf})
from Appendix~\ref{Scalar-transverse decomposition of charges}),
\begin{align}
\hat{P}_i^\star ={}&
	\hat{P}_i^{\scr T}
	+
	\int\! d^{D-1}x \,
	\Bigl(
	\hat{K}_2^\dag \hat{A}_i^{\scr L} 
	+ \hat{A}_i^{\scr L} \hat{K}_2
	- \hat{K}_1^\dag \partial_i \hat{A}_0 
	- \partial_i \hat{A}_0 \hat{K}_1
	\Bigr)
	\, ,
\label{hatPi star}
\\
\hat{M}_{ij}^\star ={}&
	\hat{M}_{ij}^{\scr T}
	+
	\int\! d^{D-1}x \,
	\Bigl(
	2 \hat{K}_1^\dag x_{[i} \partial_{j]} \hat{A}_0
	+ 2 x_{[i} \partial_{j]} \hat{A}_0 \hat{K}_1
	- 2\hat{K}_2^\dag x_{[i}  \hat{A}_{j]}^{\scr L}
	- 2 x_{[i} \hat{A}_{j]}^{\scr L} \hat{K}_2
	\Bigr)
	\, ,
\label{hatMij star}
\\
\hat{Q}^\star ={}&
	\hat{Q}^{\scr T}
	+
	\frac{1}{a}
	\int\! d^{D-1}x\, \biggl[ 
		- \frac{a^{4-D} }{2} \hat{\Psi}_2 \nabla^{-2} \hat{\Psi}_2
		- \frac{a^{4-D}}{2} \xi \hat{\Psi}_1 \hat{\Psi}_1
		- \hat{K}_2^\dag \hat{A}_0 
		- \hat{A}_0 \hat{K}_2
\nonumber \\
&
		+ \hat{K}_1^\dag \partial_i \hat{A}_i^{\scr L} 
		+ \partial_i \hat{A}_i^{\scr L} \hat{K}_1
	+ \mathcal{H} x_i \Bigl( 
		\hat{K}_2^\dag \hat{A}_i^{\scr L}
		+ \hat{A}_i^{\scr L} \hat{K}_2
		+ \partial_i\hat{K}_1^\dag \hat{A}_0
		+ \hat{A}_0 \partial_i \hat{K}_1
		\Bigr) \biggr]
		\, ,
\label{hatQ star}
\\
\hat{K}_i^\star ={}&
	\hat{K}_i^{\scr T}
	+
	\frac{1}{a}
	\int\! d^{D-1}x \, \Biggl[
		a^{4-D} \hat{K}_2^\dag \nabla^{-2} \hat{\Pi}_i^{\scr T}
		+ a^{4-D} \hat{\Pi}_i^{\scr T} \nabla^{-2} \hat{K}_2
		- \hat{K}_1^\dag \hat{A}_i^{\scr T}
		- \hat{A}_i^{\scr T}  \hat{K}_1
\nonumber \\
&
		+ (D\!-\!3) \mathcal{H} \Bigl( 
			\hat{K}_2^\dag \nabla^{-2} \hat{A}_i^{\scr T} 
			+ \hat{A}_i^{\scr T} \nabla^{-2} \hat{K}_2 
			\Bigr)
		+ \frac{ a^{4-D} }{2} x_i \Bigl( \hat{\Psi}_2 \nabla^{-2} \hat{\Psi}_2
			+ \xi \hat{\Psi}_1 \hat{\Psi}_1 \Bigr)
\nonumber \\
&
		+ x_i \biggl( 
			\partial_j \hat{K}_1^\dag \hat{A}_j^{\scr L}
			+ \hat{A}_j^{\scr L} \partial_j \hat{K}_1
			+ \hat{K}_2^\dag \hat{A}_0
			+ \hat{A}_0 \hat{K}_2
			+ (D\!-\!1) \mathcal{H} \Bigl( \hat{K}_1^\dag \hat{A}_0 
				+ \hat{A}_0 \hat{K}_1 \Bigr)
		\biggr)
\nonumber \\
&
	+
	\frac{a}{2H}
	\biggl( \frac{1}{a^2} \!-\! 1 \!-\! H^2 x_j x_j \biggr) 
	\Bigl(
	\hat{K}_1^\dag \partial_i \hat{A}_0
	+ \partial_i \hat{A}_0 \hat{K}_1
	- \hat{K}_2^\dag \hat{A}_i^{\scr L}
	- \hat{A}_i^{\scr L} \hat{K}_2
	\Bigr)
\nonumber \\
&
		+ \mathcal{H} x_i x_j 
		\Bigl( 
		\hat{K}_1^\dag \partial_j \hat{A}_0 
		+ \partial_j \hat{A}_0 \hat{K}_1
		- \hat{K}_2^\dag \hat{A}_j^{\scr L}
		- \hat{A}_j^{\scr L} \hat{K}_2
		\Bigr)
	\Biggr] \, ,
\label{hatKi star}
\end{align}
where the transverse parts are given in~(\ref{PiT}--\ref{KiT}).
The next step is requiring that all the generators commute with~$\hat{\mathcal{K}}$
 (modulo~$\hat{\mathcal{K}}$ itself). 
This requirement has two interpretations. The first is that it makes the gauge-fixed
generators into observables; the second is that it puts a condition on the non-Hermitian constraint
operator to be consistent with de Sitter symmetries. 
For spatial translations and spatial rotations this is accomplished by taking
coefficients as in~(\ref{general ansatz for K}),
\begin{equation}
\bigl[ \hat{\mathcal{K}}(\vec{k}) , \hat{P}_i^\star \bigr] =
	k_i \hat{\mathcal{K}}(\vec{k}) \, ,
\quad \qquad
\bigl[ \hat{\mathcal{K}}(\vec{k}) , \hat{M}_{ij}^\star \bigr] =
	i e^{i\theta(\vec{k})} 
	\Bigl( k_i \frac{\partial}{\partial k_j} \!-\! k_j \frac{\partial}{\partial k_i} \Bigr) 
		\Bigl( e^{-i\theta(\vec{k})}  \hat{\mathcal{K}}(\vec{k}) \Bigr) 
		\, .
\end{equation}
For dilations this is no longer the case,
\begin{align}
\bigl[ \hat{\mathcal{K}}(\vec{k}) , \hat{Q}^\star \bigr] ={}&
	\Bigl( 1 + 2 \, {\rm sh}^2[\rho(k)] \Bigr) \Bigl( k - Hk \frac{\partial \varphi(k)}{\partial k} \Bigr) \hat{\mathcal{K}}(\vec{k})
	+ e^{i\theta(\vec{k})} i H k_i \frac{\partial}{\partial k_i} \Bigl( e^{-i \theta(\vec{k}) } \hat{\mathcal{K}}(\vec{k}) \Bigr)
\nonumber \\
&	\hspace{-2cm}
	+ \frac{(D\!+\!1) i H}{2} \hat{\mathcal{K}}(\vec{k})
	- e^{i \theta(\vec{k}) + i \theta(-\vec{k})}
	\biggl[
	{\rm sh}[2\rho(k)] \Bigl( k \!-\! Hk \frac{\partial \varphi(k)}{\partial k} \Bigr)
	+ i H k \frac{\partial \rho(k)}{\partial k}
	\biggr]
	\hat{\mathcal{K}}^\dag(-\vec{k})
	\, ,
\end{align}
as the conjugate of the non-Hermitian constraint operator appears on the right hand side.
Requiring that its coefficient vanishes selects two options:
\begin{align}
&
\text{\tt option 1:} \qquad
	\rho(k) = 0 \, ,
\label{regular vacuum}
\\
&
\text{\tt option 2:} \qquad
	\rho(k) = \rho = {\rm const.} \, ,
	\qquad
	\varphi(k) = \frac{k}{H}
	\, .
\label{alpha vacuum}
\end{align}
The second option does not have a well defined flat space limit, as the Hubble rate appears in the 
denominator of the phase. This option would lead to 
the~$\alpha$-vacuum-equivalent for photons.
We do not consider this option in the remainder of the paper.
Rather, we consider only the first option~(\ref{regular vacuum}), which 
has a regular flat space limit, consistent with the choice~(\ref{transverse cond on state})
for the transverse sector.
Taking~$\rho\!=\!0$ allows one to also absorb 
phase~$\varphi(k)$ into~$\theta(\vec{k})$
since it becomes redundant; effectively we are taking~$\varphi(k)\!=\!0$. 
Moreover, we can dispense with the irrelevant phase altogether,~\footnote{
The phase~$\theta(\vec{k})$ does not affect  the
propagator. An interested reader can
easily reintroduce it to expressions~(\ref{K B final}--\ref{Ki star final}) by 
taking~$\hat{\mathcal{K}}(\vec{k}) \!\to\! e^{-i\theta(\vec{k})} \hat{\mathcal{K}}(\vec{k}) $
and~$\hat{\mathcal{B}}(\vec{k}) \!\to\! e^{-i\theta(\vec{k})} \hat{\mathcal{B}}(\vec{k}) $.
} 
and simply take,
\begin{equation}
\hat{\mathcal{K}}(\vec{k}) = \hat{b}_{\scr P}(\vec{k}) \, ,
\qquad \qquad
\hat{\mathcal{B}}(\vec{k}) = \hat{b}_{\scr H}(\vec{k}) \, .
\label{K B final}
\end{equation}
Computing the remaining commutator with
the generator of spatial special conformal transformations,
\begin{align}
\bigl[ \hat{\mathcal{K}}(\vec{k}) , \hat{K}_i^\star \bigr]
	={}&
	H k_j
		\frac{\partial^2 \hat{\mathcal{K}}(\vec{k}) }{\partial k_i\partial k_j} 
	- \frac{ H k_i }{ 2 }
		\frac{\partial^2 \hat{\mathcal{K}}(\vec{k}) }{\partial k_j \partial k_j} 
	- i k \biggl( 1+ \frac{(D\!+\!1)i H }{2k}  \biggr)
		\frac{\partial \hat{\mathcal{K}}(\vec{k})}{\partial k_i} 
\nonumber \\
&
	- \frac{ik_i}{2k}
		\biggl( 3 - \frac{(D\!-\!3)(D\!+\!1) iH }{4k} \biggr) \hat{\mathcal{K}}(\vec{k})
		\, ,
\end{align}
reveals it to be consistent with that requirement.
The generators take the form,
\begin{align}
\hat{P}_i^\star ={}&
	\hat{P}_i^{\scr T}
	+
	\int\! d^{D-1}k \, k_i \Bigl(
		\hat{\mathcal{K}}^\dag(\vec{k}) \hat{\mathcal{B}}(\vec{k})
		+
		\hat{\mathcal{B}}^\dag(\vec{k}) \hat{\mathcal{K}}(\vec{k})
		\Bigr)
		\, ,
\label{Pi star final}
\\
\hat{M}_{ij}^\star ={}&
	\hat{M}_{ij}^{\scr T}
	-
	\int\! d^{D-1}k \, 
	\biggl[
		\hat{\mathcal{B}}^\dag(\vec{k}) 
		\Bigl( ik_i \frac{\partial}{\partial k_j} \!-\! ik_j \frac{\partial}{\partial k_i} \Bigr)
		\hat{\mathcal{K}}(\vec{k})
\nonumber \\
&	\hspace{5cm}
	+ \hat{\mathcal{K}}^\dag(\vec{k}) 
		\Bigl( ik_i \frac{\partial}{\partial k_j} \!-\! ik_j \frac{\partial}{\partial k_i} \Bigr)
		\hat{\mathcal{B}}(\vec{k})
	\biggr]
	\, ,
\label{Mij star final}
\\
\hat{Q}^\star ={}&
	\hat{Q}^{\scr T}
	-
	\int\! d^{D-1}k \, 
	\Biggl[
	k \Bigl( 1 \!+\! \frac{ (D\!+\!1) i H }{2k} \Bigr) \hat{\mathcal{B}}^\dag(\vec{k}) \hat{\mathcal{K}}(\vec{k})
	+ k \Bigl( 1 \!-\! \frac{ (D\!+\!1) i H }{2k} \Bigr) \hat{\mathcal{K}}^\dag(\vec{k}) \hat{\mathcal{B}}(\vec{k})
\nonumber \\
&	\hspace{1.1cm}
	-
	\frac{(1\!-\!\xi)k}{2} \hat{\mathcal{K}}^\dag(\vec{k}) \hat{\mathcal{K}}(\vec{k})
	+ 
	i H k_i \biggl( 
	\hat{\mathcal{B}}^\dag(\vec{k}) \frac{\partial \hat{\mathcal{K}}(\vec{k}) }{ \partial k_i } 
	- \frac{\partial \hat{\mathcal{K}}^\dag(\vec{k}) }{ \partial k_i } \hat{\mathcal{B}}(\vec{k})
			\biggr)
	\Biggr]
	\, ,
\label{Q star final}
\\
\hat{K}_i^\star ={}&
	\hat{K}_i^{\scr T}
	+ \int\! d^{D-1}k \, \Biggl[
	i \hat{\mathcal{K}}^\dag (\vec{k})  \hat{\mathcal{E}}_i(\vec k) 
	- i \hat{\mathcal{E}}_i^\dag (\vec{k})  \hat{\mathcal{K}}(\vec k)
\nonumber \\
&	\hspace{-0.7cm}
	+\frac{i(1\!-\!\xi)}{4} k
		\biggl(
		\frac{ \partial \hat{\mathcal{K}}^\dag(\vec{k}) }{ \partial k_i } \hat{\mathcal{K}}(\vec{k})
		-
		\hat{\mathcal{K}}^\dag(\vec{k}) \frac{ \partial \hat{\mathcal{K}} (\vec{k}) }{ \partial k_i }
		\biggr)
	+ 
	i k \Bigl(
		1 \!+\! \frac{(D\!+\!1) iH}{2k} 
		\Bigr)
		\hat{\mathcal{B}}^\dag(\vec{k}) \frac{\partial \hat{\mathcal{K}}(\vec{k}) }{\partial k_i}
\nonumber \\
&	\hspace{-0.7cm}
	-
	i k \Bigl(
		1 \!-\! \frac{(D\!+\!1) iH}{2k}                                                                                                                                                                                                                                                                                                                                                                                                                                                                                                                                                                                                                                                                                                                                                                                                                                                                                                                                                                                                                                                                                                                                                                                                                                                                                                                                                                                                                                                                                                                                                                                                                                                                              
		\Bigr)
		\frac{\partial \hat{\mathcal{K}}^\dag (\vec{k}) }{ \partial k_i } \hat{\mathcal{B}}(\vec{k})
	+
	\frac{ik_i}{2k} \Bigl(
		3 \!-\! \frac{ i(D\!-\!3) (D\!+\!1) H }{4k} 
		\Bigr)
		\hat{\mathcal{B}}^\dag(\vec{k}) \hat{\mathcal{K}}(\vec{k})
\nonumber \\
&	\hspace{-0.7cm}
	-
	\frac{ i k_i }{ 2 k } \Bigl(
		3 \!+\! \frac{i (D\!-\!3)(D\!+\!1) H}{4k} 
		\Bigr)
		\hat{\mathcal{K}}^\dag(\vec{k}) \hat{\mathcal{B}}(\vec{k}) 
	- H k_j 
		\biggl( 
		\frac{\partial^2 \hat{\mathcal{K}}^\dag(\vec{k}) }{ \partial k_i \partial k_j} \hat{\mathcal{B}}(\vec{k}) 
		+
		\hat{\mathcal{B}}^\dag(\vec{k}) \frac{\partial^2 \hat{\mathcal{K}}(\vec{k}) }{ \partial k_i \partial k_j} 
		\biggr)
\nonumber \\
&	\hspace{-0.7cm}
	+ \frac{H k_i}{2} 
		\biggl(
		\frac{\partial^2 \hat{\mathcal{K}}^\dag(\vec{k}) }{ \partial k_j \partial k_j}
		 \hat{\mathcal{B}}(\vec{k})
		+
		\hat{\mathcal{B}}^\dag(\vec{k})
		\frac{ \partial^2 \hat{\mathcal{K}}(\vec{k}) }{ \partial k_j \partial k_j}
		\biggr)
		\Biggr]
		\, .
\label{Ki star final}
\end{align}
Having ensured that the subsidiary non-Hermitian constraint operator is consistent with de Sitter invariant
dynamics, we may finally define a de Sitter invariant state. 
 This is implemented by requiring the state to be annihilated by all the 
gauge-fixed de Sitter symmetry generators~(\ref{Pi star final})--(\ref{Ki star final}). 
Given the subsidiary condition~(\ref{momentum subsidiary}),
this is possible only if the state satisfies,
\begin{equation}
\hat{\mathcal{B}}(\vec{k}) \bigl| \Omega \bigr\rangle
	=
	0
	\, ,
	\qquad \quad
\bigl\langle \Omega \bigr| \hat{\mathcal{B}}^\dag(\vec{k})
	=
	0
	\, ,
	\qquad \quad
	\forall \vec{k}
	\, ,
\label{second scalar cond on state}
\end{equation}
Together with~(\ref{regular vacuum}) these completely define
the state,
which serves as the vacuum for constructing the indefinite inner product space of 
states~\cite{Glavan:2022a}.

\subsection{Summary: momentum space de Sitter invariant state}
\label{subsec: Momentum space de Sitter invariant state}

The most important result of this rather technical section is the momentum space
construction of a de Sitter invariant quantum state. Such construction is considerably simpler
than in position space since the spatial momentum modes do not couple at the linear level.
Here we briefly summarize the results relevant for subsequent sections.

\medskip

The physical properties of the system are described by the gauge-invariant action~(\ref{action}),
which is invariant under de Sitter transformations from Sec.~\ref{subsec: De Sitter symmetries}.
The associated generators~(\ref{Pi}--\ref{Ki}) of these transformations are the conserved 
Noether charges that characterize
the state. Requiring that all expectation values of these charges and their products vanish
defines a class of physically de Sitter invariant states, and puts conditions on the 
transverse sector of the state only. The minimum energy state in this class is defined by,
\begin{equation}
\hat{b}_{\scr T}(\sigma, \vec{k}) \bigl| \Omega \bigr\rangle = 0
 \, ,
 \qquad \qquad
 \forall \vec{k}, \sigma \, ,
\label{transverse cond on state 2}
\end{equation}
which is the state we consider as the vacuum.
The remainder of the transverse space of states is then constructed as a Fock space.
The gauge fixed dynamics is given by the gauge fixed action~(\ref{gauge fixed action}), which 
is chosen to preserve all the de Sitter symmetries. Accordingly, there are conserved 
Noether charges~(\ref{Pi star})--(\ref{Ki star}) that 
generate the gauge fixed symmetries. Requiring that the full state, that includes both 
the transverse and scalar sectors, is de Sitter symmetric requires (i) that the non-Hermitian constraint 
operator~(\ref{general ansatz for K}) commutes with all the gauge fixed generators, which 
fixes~$\hat{\mathcal{K}}(\vec{k}) \!=\! \hat{b}_{\scr P}(\vec{k}) $,
%
%
and (ii) that the state is an eigenstate of all the gauge fixed de Sitter generators mode-by-mode,
which fixes,~\footnote{This is true up to the freedom in~(\ref{alpha vacuum}), 
defining the $\alpha$-vacuum states of the scalar sector.}
\begin{equation} 
 \hat{b}_{\scr P}(\vec{k}) \bigl| \Omega \bigr\rangle=0
 \,,\qquad \quad
 \hat{b}_{\scr H}(\vec{k}) \bigl| \Omega \bigr\rangle=0
 \,,
  \qquad \quad
 \forall \vec{k}
	\, .
\label{scalar cond on state 2}
\end{equation}
The scalar sector space of states is then necessarily an indefinite inner product space, and is
constructed by acting repeatedly
with $\hat{b}_{\scr P}^\dag$ and~$\hat{b}_{\scr H}^\dag$ on~$\bigl| \Omega \bigr\rangle$.
Conditions~(\ref{transverse cond on state 2}) and~(\ref{scalar cond on state 2}) fully define the state
whose two-point functions we compute in the following section.

\section{Two-point function}
\label{sec: Two-point function}

The basic building blocks of nonequilibrium perturbation theory for interacting
vector fields in de Sitter are the various two-point functions determined in the free theory.
In this section we first briefly introduce these two-point functions and the properties they must satisfy.
We then proceed to compute the two-point functions by evaluating the integrals
over the mode functions obtained in Sec.~\ref{sec: Field operator dynamics}.
Rewriting the results in a covariant form reveals the existence of
 {\it one de Sitter breaking structure function}, which is the principal result of
this work. The section concludes by examining particular limits of the two-point functions
and comparing them with known results.

\subsection{Generalities}
\label{subsec: Generalities}

The positive-frequency Wightman two-point function of the photon is defined as an expectation
value of an unordered product of two vector field operators,
\begin{equation}
i \bigl[ \tensor*[_\mu^{\scr - \!}]{\Delta}{_\nu^{\scr \! +}} \bigr](x;x')
	=
	\bigl\langle \Omega \bigr| \hat{A}_\mu(x) \hat{A}_\nu(x') \bigl| \Omega \bigr\rangle \, .
\label{Wightman def}
\end{equation}
We compute this two-point function using the mode functions obtained in Sec.~\ref{sec: Field operator dynamics}.
Its complex conjugate defines the negative-frequency Wightman 
function,~$i \bigl[ \tensor*[_\mu^{\scr + \!}]{\Delta}{_\nu^{\scr \! -}} \bigr](x;x') \!=\! 
	\bigl\{ i \bigl[ \tensor*[_\mu^{\scr - \!}]{\Delta}{_\mu^{\scr \! +}} \bigr](x;x') \bigr\}^*$,
and the two together define the Feynman propagator,
\begin{equation}
i \bigl[ \tensor*[_\mu^{\scr + \!}]{\Delta}{_\nu^{\scr \! +}} \bigr](x;x')
	=
	\theta(\eta\!-\!\eta') \, i \bigl[ \tensor*[_\mu^{\scr - \!}]{\Delta}{_\nu^{\scr \! +}} \bigr](x;x')
	+
	\theta(\eta'\!-\!\eta)
	\, i \bigl[ \tensor*[_\mu^{\scr + \!}]{\Delta}{_\nu^{\scr \! -}} \bigr](x;x')	 
	\, .
\label{Feynman def}
\end{equation}
%
%
%
and its conjugate,~$i \bigl[ \tensor*[_\mu^{\scr -\!}]{\Delta}{_\nu^{\scr \! -}} \bigr](x;x')
=\bigl\{i \bigl[ \tensor*[_\mu^{\scr + \!}]{\Delta}{_\nu^{\scr \! +}} \bigr](x;x')\bigr\}^*$,
known as the Dyson propagator.

The field operators in the definitions of the two-point functions above satisfy equations
of motion~(\ref{position eom 1}--\ref{position eom 4}), that can be written in a more familiar covariant form,
\begin{equation}
{\mathcal{D}_\mu}^\nu \, \hat{A}_\nu(x) = 0 \, ,
\qquad \qquad
\mathcal{D}_{\mu\nu}
	= \nabla_\rho \nabla^\rho g_{\mu\nu}
	- \Bigl( 1 \!-\! \frac{1}{\xi} \Bigr) \nabla_\mu \nabla_\nu
	- R_{\mu\nu} \, .
\end{equation}
These are the equations of motion inherited by the Wightman function~(\ref{Wightman def}),
\begin{equation}
{\mathcal{D}_\mu}^\rho \, i \bigl[ \tensor*[_\rho^{\scr - \!}]{\Delta}{_\nu^{\scr \! +}} \bigr](x;x') = 0 \, ,
\qquad \qquad
{\mathcal{D}'_\nu}^\sigma \, i \bigl[ \tensor*[_\mu^{\scr - \!}]{\Delta}{_\sigma^{\scr \! +}} \bigr](x;x') = 0 \, ,
\end{equation}
while the Feynman propagator~(\ref{Feynman def}) picks up a delta function source on the account of the 
time-ordering in its definition and the canonical commutation relations~(\ref{canonical commutators}),
\begin{equation}
{\mathcal{D}_\mu}^\rho \, i \bigl[ \tensor*[_\rho^{\scr + \!}]{\Delta}{_\nu^{\scr \! +}} \bigr](x;x') 
	= g_{\mu\nu} \frac{i \delta^D(x\!-\!x')}{ \sqrt{-g} }
	\, ,
\qquad
{\mathcal{D}'_\nu}^\sigma \, i \bigl[ \tensor*[_\mu^{\scr + \!}]{\Delta}{_\sigma^{\scr \! +}} \bigr](x;x') 
		= g_{\mu\nu} \frac{i \delta^D(x\!-\!x')}{ \sqrt{-g} } \, ,
\label{Feynman EOM}
\end{equation}
The two-point functions we construct via the sum-over-modes ultimately satisfy these 
equations of motion.

 In addition to satisfying the equations of motion, the photon two-point functions
have to satisfy identities stemming from gauge invariance of the theory. These can be understood as the
two-point functions of Hermitian constraints~(\ref{constraint two-point functions}), 
which have to vanish according to the principles of
canonical quantization. 
We can express them directly in terms of the two-point functions of vector potential fields
since the two Hermitian constraints are given 
by~$\nabla^\mu \hat{A}_\mu \!=\! \xi a^{2-D} \hat{\Pi}_0$
and~$\partial_i \hat{F}_{0i} \!=\! a^{4-D} \partial_i \hat{\Pi}_i$.
The subsidiary conditions thus take the following form for the Wightman function,
\begin{subequations}
\begin{align}
\nabla^\mu \nabla'^\nu i \bigl[ \tensor*[_\mu^{\scr - \! }]{\Delta}{_\nu^{\scr \! +} } \bigr](x;x') ={}& 0 \, ,
\\
\bigl( 2 g^{ij} \delta_{[i}^\mu \partial_{0]} \partial_j \bigr) \nabla'^\nu
	i \bigl[ \tensor*[_\mu^{\scr - \! }]{\Delta}{_\nu^{\scr \! +} } \bigr](x;x') ={}& 0 \, ,
\\
\nabla^\mu \bigl( 2 g'^{kl} \delta_{[k}^\nu \partial'_{0]} \partial'_l \bigr)
	i \bigl[ \tensor*[_\mu^{\scr - \! }]{\Delta}{_\nu^{\scr \! +} } \bigr](x;x') ={}& 0 \, ,
\\
\bigl( 2 g^{ij} \delta_{[i}^\mu \partial_{0]} \partial_j \bigr)
	\bigl( 2 g'^{kl} \delta_{[k}^\nu \partial'_{0]} \partial'_l \bigr)
	i \bigl[ \tensor*[_\mu^{\scr - \! }]{\Delta}{_\nu^{\scr \! +} } \bigr](x;x') ={}& 0 \, .
\end{align}
\end{subequations}
while for the Feynman propagator they also receive a local contribution due to
the time-ordered product of operators in its definition,
\begin{subequations}
\begin{align}
\nabla^\mu \nabla'^\nu i \bigl[ \tensor*[_\mu^{\scr + \! }]{\Delta}{_\nu^{\scr \! +} } \bigr](x;x') ={}& 
	-\xi \frac{i \delta^{D}(x\!-\!x')}{\sqrt{-g}} \, ,
\\
\bigl( 2 g^{ij} \delta_{[i}^\mu \partial_{0]} \partial_j \bigr) \nabla'^\nu
	i \bigl[ \tensor*[_\mu^{\scr + \! }]{\Delta}{_\nu^{\scr \! +} } \bigr](x;x') ={}& 0 \, ,
\\
\nabla^\mu \bigl( 2 g'^{kl} \delta_{[k}^\nu \partial'_{0]} \partial'_l \bigr)
	i \bigl[ \tensor*[_\mu^{\scr + \! }]{\Delta}{_\nu^{\scr \! +} } \bigr](x;x') ={}& 0 \, ,
\\
\bigl( 2 g^{ij} \delta_{[i}^\mu \partial_{0]} \partial_j \bigr)
	\bigl( 2 g'^{kl} \delta_{[k}^\nu \partial'_{0]} \partial'_l \bigr)
	i \bigl[ \tensor*[_\mu^{\scr + \! }]{\Delta}{_\nu^{\scr \! +} } \bigr](x;x') ={}& 
	- \nabla^2 \frac{i \delta^{D}(x\!-\!x')}{\sqrt{-g}} \, .
\end{align}
\label{Feynman subsidiary}%
\end{subequations}
%

\subsection{Sum over modes}
\label{subsec: Sum over modes}

Given the mode decompositions~(\ref{Fourier transforms}),
and the solutions for the momentum space field operators~(\ref{AT final solution}),~(\ref{A 0 final solution}), 
and~(\ref{A L final solution}), and the definition of the physical state~(\ref{transverse cond on state}) and~(\ref{momentum subsidiary}),
components of the photon
two-point function can be expressed as
the following integrals over the mode functions,
\begin{align}
\MoveEqLeft[4]
i \bigl[ \tensor*[_0^{\scr - \! }]{\Delta}{_0^{\scr \! +}} \bigr] (x;x')
\label{mode sum 00}
\\
={}&
	(aa')^{-\frac{D-2}{2}} \!\! \int\! \frac{ d^{D-1} k }{ (2\pi)^{D-1} } \, e^{i \vec{k} \cdot ( \vec{x} - \vec{x}^{\,\prime} ) }
	\biggl[
	- U_{\nu}(\eta,k) v_{0}^*(\eta',k) - v_0(\eta,k) U_{\nu}^*(\eta',k)
	\biggr] \, ,
\nonumber \\
\MoveEqLeft[4]
i \bigl[ \tensor*[_0^{\scr - \! }]{\Delta}{_i^{\scr \! +}} \bigr] (x;x')
\label{mode sum 0i}
\\
	={}&
	(aa')^{-\frac{D-2}{2}} \!\! \int\! \frac{ d^{D-1} k }{ (2\pi)^{D-1} } \, e^{i \vec{k} \cdot ( \vec{x} - \vec{x}^{\,\prime} ) }
	\frac{k_i}{k} \biggl[
	U_{\nu}(\eta,k) v_{\scr L}^*(\eta',k) + v_0(\eta,k) U_{\nu+1}^*(\eta',k)
	\biggr] \, ,
\nonumber \\
\MoveEqLeft[4]
i \bigl[ \tensor*[_{i\,}^{\scr - \! }]{\Delta}{_j^{\scr \! +}} \bigr] (x;x')
	=
	(aa')^{-\frac{D-4}{2}} \!\! \int\! \frac{ d^{D-1}k }{ (2\pi)^{D-1} } \, e^{i \vec{k} \cdot ( \vec{x} - \vec{x}^{\,\prime} ) } \,
	\Bigl( \delta_{ij} \!-\! \frac{ k_i k_j }{ k^2 } \Bigr) \,
	U_{\nu}(\eta,k) U_{\nu}(\eta',k)
\label{mode sum ij}
\\
&	\hspace{-1.cm}
	- (aa')^{-\frac{D-2}{2} } \!\! \int\! \frac{ d^{D-1}k }{ (2\pi)^{D-1} } \, e^{i \vec{k} \cdot ( \vec{x} - \vec{x}^{\,\prime} ) } \,
	\frac{ k_i k_j }{ k^2 }
	\biggl[
	U_{\nu+1}(\eta,k) v_{\scr L}^*(\eta',k) + v_{\scr L}(\eta,k) U_{\nu+1}^*(\eta',k)
	\biggr]
	\, ,
\nonumber 
\end{align}
where we made use of Eqs.~(\ref{Wronskian condition}), (\ref{Wronskian-like}) 
and the~$i\varepsilon$-prescriptions are implicit in the same way as for the scalar 
propagator~(\ref{int over modes}). In the following two sections we solve these integrals
using the solutions for the mode functions found in Sec.~\ref{sec: Field operator dynamics}.
Plugging in the de Sitter space particular functions~(\ref{dS particular mode 0})
and~(\ref{dS particular mode L}) into the sum over modes expressions~(\ref{mode sum 00})--(\ref{mode sum ij}),
using recurrence relations for mode functions~(\ref{mode recurrence}),
and recognizing the scalar two-point functions~(\ref{int over modes}) produces 
expressions,~\footnote{Function~$\mathcal{F}_{\nu+1}(y)$ appearing in 
expressions~(\ref{dS 00 first}--\ref{dS ij first}) is divergent in any dimension, but
it is implied that all the derivatives acting on it, including the parametric one, are to be taken first, and 
only then 
the index set to~$\nu\!=\!(D\!-\!3)/2$, which produces a well defined finite result.}
\begin{align}
i \bigl[ \tensor*[_0^{\scr - \!}]{\Delta}{_0^{\scr \! +}} \bigr] (x;x')
	={}&
	\frac{ - \xi}{ 2 (\nu\!+\!1) H^2 } 
		\Bigl[ 
		\mathcal{H} \partial_0 + \mathcal{H}' \partial'_0 
		+ (D\!-\!2)
			\bigl( \mathcal{H}^2 \!+\! \mathcal{H}'^2 \bigr) 
		\Bigr] \mathcal{F}_{\nu}(y)
\nonumber \\
&
	+
	\Bigl( 1 \!-\! \frac{ \xi }{ \xi_s } \Bigr)
	\frac{ \nabla^2 }{ 2\nu H^2}
	\frac{ \partial }{ \partial \nu }
	\mathcal{F}_{\nu}(y)
	\, ,
\label{dS 00 first}
\\
i \bigl[ \tensor*[_0^{\scr - \!}]{\Delta}{_i^{\scr \! +}} \bigr] (x;x')
	={}&
	\frac{ \partial_i' }{2\nu H^2} 
		\Bigl[ \mathcal{H} \mathcal{F}_{\nu+1}(y)
			- \mathcal{H}' \mathcal{F}_{\nu}(y) \Bigr]
	- \Bigl( 1 \!-\! \frac{ \xi }{ \xi_s } \Bigr) \frac{\partial_i' \partial_0}{ 2 \nu H^2}
		\frac{\partial}{\partial \nu} \mathcal{F}_{\nu+1}(y)
		\, ,
\label{dS 0i first}
\\
i \bigl[ \tensor*[_{i\,}^{\scr - \!}]{\Delta}{_j^{\scr \! +}} \bigr] (x;x')
	={}&
	\frac{ \delta_{ij} \mathcal{H} \mathcal{H}'}{ H^2} \mathcal{F}_{\nu}(y)
	- \Bigl( 1 \!-\! \frac{ \xi }{ \xi_s } \Bigr) \frac{\partial_i \partial'_j }{ 2 \nu H^2}
		\frac{\partial}{\partial \nu} \mathcal{F}_{ \nu+1}(y)
\nonumber \\
&
	- \frac{1}{2\nu H^2} \frac{ \partial_i \partial'_j }{ \nabla^2 }
		\Bigl[ \mathcal{H} \partial'_0 + \mathcal{H}' \partial_0 + (D\!-\!1) \mathcal{H} \mathcal{H}' \Bigr]
			\mathcal{F}_{\nu}(y) \, .
\label{dS ij first}
\end{align}
It is advantageous to rewrite these expressions by using the identity~\cite{Glavan:2020zne},
\begin{equation}
\Bigl[ \mathcal{H}' \partial_0 + \mathcal{H} \partial_0' + (D\!-\!1) \mathcal{H} \mathcal{H}' \Bigr] f(y)
	= \frac{1}{2} \nabla^2 I[f](y) \, ,
\end{equation}
to eliminate the inverse Laplacian in the last component, and by acting some derivatives explicitly
to obtain,
\begin{align}
i \bigl[ \tensor*[_0^{\scr - \!}]{\Delta}{_0^{\scr \! +}} \bigr] (x;x')
	={}&
	\frac{ 1}{ 2 \nu H^2 } \frac{\xi}{\xi_s}
		\biggl\{
		\bigl( \mathcal{H}^2 \!+\! \mathcal{H}'^2 \bigr) \biggl[ (2\!-\!y) \frac{\partial}{\partial y} \!-\! (D\!-\!2) \biggr]
		- 4 \mathcal{H} \mathcal{H}' \frac{\partial}{\partial y}
		\biggr\} \mathcal{F}_{\nu}(y)
\nonumber \\
&	\hspace{-2.5cm}
	+
	\frac{ 1 }{ 2\nu H^2}
	\Bigl( 1 \!-\! \frac{ \xi }{ \xi_s } \Bigr)
	\biggl\{ 
	4 \bigl( \mathcal{H}^2 \!+\! \mathcal{H}'^2 \bigr) \frac{\partial}{\partial y}
	- 2 \mathcal{H} \mathcal{H}' \biggl[ 2 (2\!-\!y) \frac{\partial}{\partial y} \!-\! (D\!-\!1) \biggr]
	\biggr\}
	\frac{\partial}{\partial y} \frac{\partial}{\partial \nu} \mathcal{F}_{\nu}(y)
	\, ,
\label{dS component 00}
\\
i \bigl[ \tensor*[_0^{\scr - \!}]{\Delta}{_i^{\scr \! +}} \bigr] (x;x')
	={}&
	\frac{ \bigl( \partial'_i y \bigr) }{ 2 \nu H^2 } 
		\biggl[ \mathcal{H} \frac{\partial}{\partial y} \mathcal{F}_{\nu+1}(y)
		- \mathcal{H}' \frac{\partial}{\partial y} \mathcal{F}_{\nu}(y) \biggr]
\nonumber \\
&	\hspace{-1cm}
	- \frac{ \bigl( \partial'_i y \bigr) }{ 2 \nu H^2 }
	\Bigl( 1 \!-\! \frac{ \xi }{ \xi_s } \Bigr)
	\biggl\{ 2 \mathcal{H}' \frac{\partial}{\partial y}
		+ \mathcal{H} \biggl[ 1 \!-\! (2\!-\!y) \frac{\partial}{\partial y} \biggr] \biggr\} 
		\frac{\partial}{\partial y} \frac{\partial}{\partial\nu} \mathcal{F}_{\nu+1}(y)
		\, ,
\label{dS component 0i}
\\
i \bigl[ \tensor*[_{i\,}^{\scr - \!}]{\Delta}{_j^{\scr \! +}} \bigr] (x;x')
	={}&
	\frac{ \delta_{ij} \mathcal{H} \mathcal{H}'}{ H^2} \mathcal{F}_{\nu}(y)
	- \frac{ \partial_i \partial'_j }{2\nu H^2} \biggl[ 
		\frac{1}{2} I\big[\mathcal{F}_{\nu}\big](y)
		+ \Bigl( 1 \!-\! \frac{ \xi }{ \xi_s } \Bigr) 
			\frac{\partial}{\partial \nu} \mathcal{F}_{\nu+1}(y)
	\biggr]
	\, .
\label{dS component ij}
\end{align}
In this
way we have accomplished writing the photon two-point function in terms of scalar two-point 
functions and their derivatives.  However,
this form of the two-point function is not practical, as it is not covariant, and moreover 
its properties are not obvious. In what follows we derive a more systematic, covariant representation.

\subsection{Covariantization: de Sitter invariant ansatz}
\label{subsec: Covariantization: de Sitter invariant ansatz}

At this point there is nothing that motivates us to consider anything else than 
a de Sitter invariant ansatz,
\begin{equation}
i \bigl[ \tensor*[_\mu^{\scr - \!}]{ \Delta }{_\nu^{\scr \! +}} \bigr]^{\rm \scr dS} (x;x')
	=
	\bigl( \partial_\mu \partial'_\nu y \bigr) \, \mathcal{C}_1^{\rm \scr dS}(y)
	+
	\bigl( \partial_\mu y \bigr) \bigl( \partial'_\nu y \bigr) \, \mathcal{C}_2^{\rm \scr dS}(y) \, ,
\label{dS covariant basis}
\end{equation}
where~$\mathcal{C}_1$ and~$\mathcal{C}_2$ are the two scalar structure functions.
We have to compare this expression to~(\ref{dS component 00}--\ref{dS component ij}) 
to solve for the structure functions. Writing out the components of~(\ref{dS covariant basis})
gives,
\begin{align}
i \bigl[ \tensor*[_{i\,}^{\scr - \!}]{\Delta}{_j^{\scr \! +}} \bigr]^{\rm \scr dS}(x;x') 
	={}&
	2 \delta_{ij} \mathcal{H} \mathcal{H}' \Bigl[ I [\mathcal{C}_2^{\rm \scr dS}] - \mathcal{C}_1^{\rm \scr dS} \Bigr]
	+ \partial_i \partial'_j I^2[\mathcal{C}_2^{\rm \scr dS}]
	\, ,
\\
i \bigl[ \tensor*[_0^{\scr - \!}]{\Delta}{_i^{\scr \! +}} \bigr]^{\rm \scr dS} (x;x') 
	={}&
	(\partial'_i y) \biggl\{
		\mathcal{H} \Bigl[ \mathcal{C}_1^{\rm \scr dS} - (2\!-\!y) \mathcal{C}_2^{\rm \scr dS} \Bigr]
			+ 2 \mathcal{H}' \mathcal{C}_2^{\rm \scr dS}
		\biggr\}
		\, ,
\\
i \bigl[ \tensor*[_0^{\scr - \!}]{\Delta}{_0^{\scr \! +}} \bigr]^{\rm \scr dS} (x;x') 
	={}& \!
	2 \bigl( \mathcal{H}^2 \!+\! \mathcal{H}'^2 \bigr) 
	\Bigl[ \mathcal{C}_1^{\rm \scr dS} \!-\! (2\!-\!y) \mathcal{C}_2^{\rm \scr dS} \Bigr]
	\!+\! \mathcal{H} \mathcal{H}' \Bigl[ - (2\!-\!y) \mathcal{C}_1^{\rm \scr dS} \!+\! \bigl( 8 \!-\! 4y \!+\! y^2 \bigr) \mathcal{C}_2^{\rm \scr dS} \Bigr]
	.
\end{align}
Comparing just the~$(ij)$ component immediately yields the solution for the two structure functions,
\begin{align}
\mathcal{C}_1^{\rm \scr dS}(y)
	={}&
	\frac{1}{2 \nu H^2}
	\biggl[
	- \Bigl( \nu\!+\! \frac{1}{2} \Bigr) \mathcal{F}_{\nu}(y)
	- \Bigl( 1 \!-\! \frac{ \xi }{ \xi_s } \Bigr)
		\frac{\partial}{\partial y} \frac{\partial}{\partial \nu} \mathcal{F}_{\nu+1}(y)
	\biggr]
	\, ,
\label{C1}
\\
\mathcal{C}_2^{\rm \scr dS}(y)
	={}&
	\frac{1}{2 \nu H^2}
	\biggl[
	- \frac{1}{2} \frac{\partial}{\partial y} \mathcal{F}_{\nu}(y)
	- \Bigl( 1 \!-\! \frac{ \xi }{ \xi_s } \Bigr)
		\frac{\partial^2 }{\partial y^2} \frac{\partial}{\partial \nu} \mathcal{F}_{\nu+1}(y)
	\biggr]
	\, .
\label{C2}
\end{align}
 These two de Sitter invariant structure functions, that multiply the two de Sitter invariant
tensor structures in~(\ref{dS covariant basis}), completely account for all the previous results
on the photon two-point functions reported in the 
literature~\cite{Allen:1985wd,Tsamis:2006gj,Youssef:2010dw,Frob:2013qsa}.
Nonetheless, it is important to show that this solution correctly reproduces 
the remaining components. Identity~(\ref{contiguousF 2}) implies this is true for the~($0i$) 
component. However, a judicious comparison of the~($00$) components gives,
\begin{align}
i \bigl[ \tensor*[_0^{\scr - \!} ]{\Delta}{_0 ^{\scr \! +} } \bigr] (x;x')
	-
	i \bigl[ \tensor*[_0^{\scr - \!} ]{ \Delta }{_0 ^{\scr \! +} } \bigr]^{\rm \scr dS} (x;x')
	={}&
	- 
	\frac{ \mathcal{H}\mathcal{H}'}{2\nu H^2} 
	\frac{\xi}{\xi_s}
	\biggl[ \Bigl( \nu \! -\! \frac{D\!-\!3}{2} \Bigr) \mathcal{F}_{\nu+1}(y) \biggr]_{\nu \to \frac{D-3}{2}}
\nonumber \\
	={}&
	\xi \times
	aa' \frac{ H^{D-2} }{ (4\pi)^{\frac{D}{2}} }
		\frac{ \Gamma(D\!-\!1) }{ (D\!-\!1) \, \Gamma\bigl( \frac{D}{2} \bigr) }
		\, .
\label{de Sitter breaking term}
\end{align}
upon applying equation~(\ref{F eom}),
recurrence identities~(\ref{contiguousF 1}),~(\ref{contiguousF 2}), 
and the limit (\ref{limit of F nu+1}).
The fact that the de Sitter invariant ansatz~(\ref{dS covariant basis}) cannot reproduce all
the components of the would-be de Sitter invariant propagator means that
{\it there is no de Sitter invariant propagator, except in the limit~$\xi\!\to\!0$.}


\subsection{Covariantization: de Sitter breaking ansatz}
\label{subsec: Covariantization: de Sitter breaking ansatz}

Having discovered that there cannot be a de Sitter invariant 
solution, we make the following, more general, {\it Ansatz},~\footnote{Strictly 
speaking we could have made a more general {\it Ansatz} 
including a dependence on~$v\!=\!\ln(a/a')$, but this would lead to the same result.}
\begin{align}
i \bigl[ \tensor*[_\mu^{\scr - \!}]{\Delta}{_\nu^{\scr \! +}} \bigr] (x;x')
	={}&
	\bigl( \partial_\mu \partial'_\nu y \bigr) \, \mathcal{C}_1(y,u)
	+
	\bigl( \partial_\mu y \bigr) \bigl( \partial'_\nu y \bigr) \, \mathcal{C}_2(y,u)
\label{covariant basis}
\\
&
	+ \Bigl[ \bigl( \partial_\mu y \bigr) \bigl( \partial'_\nu u \bigr) 
		\!+\! \bigl( \partial_\mu u \bigr) \bigl( \partial'_\nu y \bigr) \Bigr] \mathcal{C}_3(y,u)
	+ \bigl( \partial_\mu u \bigr) \bigl( \partial'_\nu u \bigr) \, \mathcal{C}_4(y,u)
	\, .
\nonumber 
\end{align}
where~$u \!=\! \ln(aa')$. 
Writing out the components of the
 tensor structures in~(\ref{covariant basis}),
\begin{align}
i \bigl[ \tensor*[_{i\,}^{\scr - \!}]{\Delta}{_j^{\scr \! +}} \bigr](x;x') 
	={}&
	2 \delta_{ij} \mathcal{H} \mathcal{H}' \Bigl\{ I [\mathcal{C}_2] - \mathcal{C}_1 \Bigr\}
	+ \partial_i \partial'_j I^2[\mathcal{C}_2]
	\, ,
\\
i \bigl[ \tensor*[_0^{\scr - \!}]{\Delta}{_i^{\scr \! +}} \bigr](x;x') 
	={}&
	(\partial'_i y) \biggl\{
		\mathcal{H} \Bigl[ \mathcal{C}_1 - (2\!-\!y) \mathcal{C}_2 + \mathcal{C}_3 \Bigr]
			+ 2 \mathcal{H}' \mathcal{C}_2
		\biggr\}
		\, ,
\\
i \bigl[ \tensor*[_0^{\scr - \!}]{\Delta}{_0^{\scr \! +}} \bigr](x;x') 
	={}& \!
	2 \bigl( \mathcal{H}^2 \!+\! \mathcal{H}'^2 \bigr) \Bigl[ \mathcal{C}_1 - (2\!-\!y) \mathcal{C}_2 + \mathcal{C}_3 \Bigr]
\nonumber \\
&
	+ \mathcal{H} \mathcal{H}' \Bigl[ - (2\!-\!y) \mathcal{C}_1 \!+\! (8 \!-\! 4y \!+\! y^2) \mathcal{C}_2
		- 2 (2\!-\!y) \mathcal{C}_3 + \mathcal{C}_4 \Bigr]
	\, ,
\end{align}
results in
a form that is straightforward to compare to~(\ref{dS component 00}--\ref{dS component ij}).
Comparing first the~$(ij)$ components yields for the first two structure functions,
\begin{align}
\mathcal{C}_1(y)
	={}&
	\frac{1}{2 \nu H^2}
	\biggl[
	- \Bigl( \nu\!+\! \frac{1}{2} \Bigr) \mathcal{F}_{\nu}(y)
	- \Bigl( 1 \!-\! \frac{ \xi }{ \xi_s } \Bigr)
		\frac{\partial}{\partial y} \frac{\partial}{\partial \nu} \mathcal{F}_{\nu+1}(y)
	\biggr]
	\, ,
\label{C1}
\\
\mathcal{C}_2(y)
	={}&
	\frac{1}{2 \nu H^2}
	\biggl[
	- \frac{1}{2} \frac{\partial}{\partial y} \mathcal{F}_{\nu}(y)
	- \Bigl( 1 \!-\! \frac{ \xi }{ \xi_s } \Bigr)
		\frac{\partial^2 }{\partial y^2} \frac{\partial}{\partial \nu} \mathcal{F}_{\nu+1}(y)
	\biggr]
	\, ,
\label{C2}
\end{align}
and comparing the~$(0i)$ component yields,
\begin{equation}
\mathcal{C}_3 = 0 \, .
\label{C3}
\end{equation}
Lastly, comparing the~$(00)$ components yields an unexpected result for the last structure function,
\begin{equation}
\mathcal{C}_4 = -
	\frac{1}{2\nu H^2} 
	\frac{\xi}{\xi_s}
	\biggl[ \Bigl( \nu \! -\! \frac{D\!-\!3}{2} \Bigr) \mathcal{F}_{\nu+1}(y) \biggr]_{\nu \to \frac{D-3}{2}}
	=
	\xi \times 
	\frac{H^{D-4}}{ (4\pi)^{\frac{D}{2}} } \frac{ \Gamma(D\!-\!1) }{ (D \!-\!1) \, \Gamma\bigl( \frac{D}{2} \bigr) }
	 \, ,
\label{C4}
\end{equation}
which does not vanish! 
This is in accordance with
Sec.~\ref{subsec: Covariantization: de Sitter invariant ansatz},
where we showed that
a fully de Sitter invariant form cannot reproduce all the components obtained from mode sums.

\medskip

The form of the photon two-point function~(\ref{covariant basis}) with the scalar structure 
functions~(\ref{C1}--\ref{C4}) comprises our solution for the Wightman function. The Feynman
propagator is obtained simply by changing every~$y_{\scr -+}$ to~$y_{\scr ++}$.


\subsection{Various limits}
\label{subsec: Various limits}

\noindent {\bf Four-dimensional limit.}
In the~$D\!\to\!4$ limit we have that~$\nu\!\to\!\frac{1}{2}$, 
and~$\xi_s \!\to\! 3$,
and that the rescaled propagator functions 
reduce to,
\begin{equation}
\mathcal{F}_{\nu}(y)
	\xrightarrow{D\to4}
	\frac{ H^2}{4\pi^2 y} 
	\, ,
\qquad
\frac{\partial}{\partial y} \frac{\partial}{\partial \nu} \mathcal{F}_{\nu+1}(y)
	\xrightarrow{D\to4}
	\frac{H^2}{ (4\pi)^2 }
	\biggl[
	- \frac{3}{y}
	+ \frac{1}{4 \!-\! y}
	+ \frac{2 (6 \!-\! y)}{(4 \!-\! y)^2} \ln\Bigl( \frac{y}{4} \Bigr)
	\biggr]
		\, .
\label{F nu in 4}
\end{equation}
Therefore, the photon two-point function~(\ref{covariant basis})
 in~$D\!=\!4$ is given by,
\begin{subequations}
\begin{align}
\mathcal{C}_1(y)
	\xrightarrow{D\to4}{}&
	\frac{1}{ (4\pi)^2 }
	\biggl\{
	- \frac{ 4}{ y} 
	- \Bigl( 1 \!-\! \frac{ \xi }{ \xi_s } \Bigr)
	\biggl[
	- \frac{3}{y}
	+ \frac{1}{4 \!-\! y}
	+ \frac{2 (6 \!-\! y)}{(4 \!-\! y)^2} \ln\Bigl( \frac{y}{4} \Bigr)
	\biggr]
	\biggr\}
	\, ,
\label{C1 in 4}
\\
\mathcal{C}_2(y)
	\xrightarrow{D\to4}{}&
	\frac{1}{ (4\pi)^2 }
	\biggl\{
	\frac{ 2}{ y^2} 
	- \Bigl( 1 \!-\! \frac{ \xi }{ \xi_s } \Bigr)
	\biggl[
	\frac{3}{y^2}
	+ \frac{ (12 \!-\! y)}{ y (4 \!-\! y)^2} 
	+ \frac{ 2 (8 \!-\! y)}{(4 \!-\! y)^3} \ln\Bigl( \frac{y}{4} \Bigr)
	\biggr]
	\biggr\}
	\, ,
\label{C2 in 4}
\\
\mathcal{C}_4 
	\xrightarrow{D\to4}{}&
	 \frac{1}{ (4\pi)^2 }
	 \times
	 \frac{ 2 \xi }{ 3 } 
	 \, .
\label{C4 in 4}
\end{align}
\end{subequations}

\bigskip

\noindent {\bf Flat space limit.}
The three tensor structures from~(\ref{covariant basis}) in flat space reduce to,
\begin{equation}
\bigl( \partial_\mu \partial'_\nu y \bigr)
	\, \overset{H\to0}{\longsim}
	- 2 H^2 \eta_{\mu\nu} \, ,
\quad
\bigl( \partial_\mu y \bigr) \bigl( \partial'_\nu y \bigr)
	\, \overset{H\to0}{\longsim} \,
	- 4 H^4 \Delta x_\mu \Delta x_\nu
	\, ,
\quad
\bigl( \partial_\mu u \bigr) \bigl( \partial'_\nu u \bigr)
	\, \overset{H\to0}{\longsim} \,
	H^2 \delta_\mu^0 \delta_\nu^0
	\, ,
\label{tensor flat limit}
\end{equation}
while from the power series representation~(\ref{power series})
we infer that the propagator function reduces to the flat space scalar two-point function,
and that its relevant derivatives are also proportional to it,
\begin{equation}
\mathcal{F}_{\nu}(y) 
	\xrightarrow{H\to0}
	\frac{\Gamma\bigl( \frac{D-2}{2} \bigr) }{ 4\pi^{ \frac{D}{2} } \bigl( \Delta x^2 \bigr)^{\!\frac{D-2}{2}} }
	\, ,
\qquad
\frac{\partial}{\partial y} \frac{\partial}{\partial \nu} \mathcal{F}_{\nu+1}(y) 
	\xrightarrow{H\to0}
	- \frac{ (D\!-\!1) }{ 4 } \times
	\frac{ \Gamma\bigl( \frac{D-2}{2} \bigr) }{ 4 \pi^{ \frac{D}{2} } \bigl( \Delta x^2 \bigr)^{\!\frac{D-2}{2}}  }
	\, ,
\label{propagator flat limit}
\end{equation}
while the derivative with respect to~$y$ reduces to,
\begin{equation}
\frac{\partial}{\partial y}
	\, \overset{H\to0}{\longsim} \,
	\frac{1}{H^2} \frac{\partial}{\partial (\Delta x^2)} \, .
\label{derivative flat limit}
\end{equation}
Thus the vector two-point function in~(\ref{covariant basis})
reproduces the correct flat space limit,
\begin{equation}
i \bigl[ \tensor*[_\mu^{\scr - \!}]{\Delta}{_\nu^{\scr \! +}} \bigr] (x;x')
	\xrightarrow{H\to0}
	\biggl[
	\eta_{\mu\nu} 
	-
	\frac{(1\!-\!\xi)}{2}
	\biggl(\!
	\eta_{\mu\nu}
	\!-\!
	(D\!-\!2) \frac{ \Delta x_\mu \Delta x_\nu }{ \Delta x^2 }
	\biggr)
	\biggr]
	\frac{\Gamma\bigl( \frac{D-2}{2} \bigr) }{ 4\pi^{ \frac{D}{2} } \bigl( \Delta x^2 \bigr)^{\!\frac{D-2}{2}} }
\,,
\label{Minkowski limit vector propagator 1}
\end{equation}
that can be also written in a more familiar form as, 
\begin{equation}
i \bigl[ \tensor*[_\mu^{\scr - \!}]{\Delta}{_\nu^{\scr \! +}} \bigr] (x;x')
	\xrightarrow{H\to0}
	\biggl[
	\eta_{\mu\nu} 
	\!-\!
	(1\!-\!\xi)\frac{\partial_\mu \partial_\nu}{\partial^2}
	\biggr]	
	\frac{\Gamma\bigl( \frac{D-2}{2} \bigr) }{ 4\pi^{ \frac{D}{2} } \bigl( \Delta x^2 \bigr)^{\!\frac{D-2}{2}} }
	\, ,
\label{Minkowski limit vector propagator 2}
\end{equation}
as a projector acting on the scalar two-point function.

\section{Simple observables}
\label{sec: Simple observables}

The two-point functions worked out in the preceding section are appropriate to
use in loop computations. Here we demonstrate their consistency by
computing two simple observables at leading orders. The first is the field strength 
correlator whose leading contribution comes in at tree level. The second is the expectation
value of the energy momentum tensor which is an example of a simple one-loop observable 
where a single two-point function forms the loop.

\subsection{Field strength correlator}
\label{subsec: Field strength correlator}

The tree-level correlator of the field stress tensor is conveniently expressed as
antisymmetrized derivatives acting on the Wightman function~(\ref{Wightman def}),
\begin{equation}
\bigl\langle \Omega \bigr| \hat{F}_{\mu\nu}(x) \hat{F}_{\rho\sigma}(x') \bigl| \Omega \bigr\rangle
	= 4 \bigl( \delta^\alpha_{[\mu} \partial_{\nu]} \bigr) 
		\bigl( \delta^\beta_{[\rho} \partial'_{\sigma]} \bigr)
		i \bigl[ \tensor*[_\alpha^{\scr - \!}]{\Delta}{_\beta^{\scr \! +}} \bigr](x;x')
		\, .
\label{F correlator def}
\end{equation}
It clearly constitutes an observable since the field strength tensor is gauge independent.
It is immediately clear that the de Sitter breaking part does not contribute to this quantity
on the account of anti-symmetrized derivatives. Acting the derivatives above onto the
covariantized representation of the two-point function~(\ref{covariant basis}) gives,
\begin{equation}
\bigl\langle \Omega \bigr| \hat{F}_{\mu\nu}(x) \hat{F}_{\rho\sigma}(x') \bigl| \Omega \bigr\rangle
	= - \frac{ 2 }{ H^2 } \biggl[ \bigl( \partial_\mu \partial'_{[\rho} y \bigr) \bigl( \partial'_{\sigma]} \partial_\nu y \bigr)
		\frac{\partial \mathcal{F}_{\nu} }{ \partial y}
	+
	\bigl( \partial_{[\mu} y \bigr) \bigl( \partial_{\nu]} \partial'_{[\sigma} y \bigr)
		\bigl( \partial'_{\rho]} y \bigr) \frac{\partial^2 \mathcal{F}_{\nu} }{ \partial y^2 }
		\biggr]
		\, .
\label{F corr dS}
\end{equation}
The fact that the correlator does not depend on the gauge-fixing parameter~$\xi$ reflects that 
this correlator is indeed an observable. The de Sitter invariance of the correlator reflects the 
fact that the state is physically de Sitter invariant.~\footnote{The phenomenon when the
one-point or two-point function 
breaks the symmetry of the background, but the corresponding energy-momentum tensor 
does not, is known as {\it symmetry non-inheritance},
one example of which is discussed in Ref.~\cite{Glavan:2019yfc} 
(see also Refs.~\cite{Herdeiro:2014goa,Cvitan:2015aha})
in which it was shown that, even though the scalar one-point function of a massless scalar field
sourced by a point charge breaks dilatation symmetry, the corresponding 
energy-momentum tensor does not.}
In four spacetime dimensions it also becomes obvious that the physical
photon couples conformally to gravity. In fact, using~(\ref{F nu in 4}) 
it follows that in~$D\!\to\!4$ 
the correlator reduces to the flat space result, 
\begin{align}
\bigl\langle \Omega \bigr| \hat{F}_{\mu\nu}(x) \hat{F}_{\rho\sigma}(x') \bigl| \Omega \bigr\rangle
	\xrightarrow{D\to4}{}&
	\frac{2}{\pi^2}\partial_{\mu]}\partial'_{[\rho}
	 \biggl[ \frac{\Delta x_{\sigma]}\Delta x_{[\nu}}{\bigl(\Delta x^2 \bigr)^{\!2} } \biggr]
\nonumber\\
	={}&
	\frac{2}{\pi^2 \bigl(\Delta x^2 \bigr)^{\!2} }
	\biggl[
	\eta_{\mu[\rho} \eta_{\sigma]\nu} 
	- 4 \eta_{\alpha [\mu} \eta_{\nu] [\sigma } \eta_{\rho] \beta}
		\frac{ \Delta x^\alpha \Delta x^\beta }{ \Delta x^2}
	\biggr]
	\, .
\label{FF off coincident correlator}
\end{align}
%

\subsection{Energy-momentum tensor}
\label{subsec: Energy-momentum tensor}

The energy-momentum tensor of the photon field is defined as a variation of the 
action with respect to the metric tensor. Two definitions are thus possible since we
can either consider the gauge-invariant action~(\ref{action}), 
or the gauge-fixed action~(\ref{gauge fixed action})
 that
in addition contains the gauge-fixing part~(\ref{intro gauge}). The two definitions 
give the same answer at the level of expectation value, as they should.

\medskip
\noindent{\bf Gauge-invariant energy-momentum tensor.}
The energy-momentum tensor defined from the gauge-invariant action,
\begin{equation}
T_{\mu\nu} = \frac{-2}{\sqrt{-g}} \frac{\delta S}{\delta g^{\mu\nu}}
	= \Bigl( \delta_\mu^\rho \delta_\nu^\sigma - \frac{1}{4} g_{\mu\nu} g^{\rho\sigma} \Bigr) 
		g^{\alpha\beta} F_{\rho\alpha} F_{\sigma\beta}
		\, ,
\label{T def}
\end{equation}
is manifestly gauge-independent as it depends on the field strength tensor only.
All of its components consist only of transverse fields and of the secondary first-class
constraint~$\Psi_2$~\cite{Glavan:2022a}.
This is why we need not
 worry about operator ordering of constraints 
when constructing the operator associated with the observable.
It is defined by Weyl-ordering the
products, and consequently its expectation value is,
\begin{equation}
\bigl\langle \Omega \bigr| \hat{T}_{\mu\nu}(x) \bigl| \Omega \bigr\rangle
	=
	\Bigl( \delta_{(\mu}^\rho \delta_{\nu)}^\sigma - \frac{1}{4} g_{\mu\nu} g^{\rho\sigma} \Bigr) 
		g^{\alpha\beta} 
		\times \frac{1}{2} \bigl\langle \Omega \bigr| \bigl\{ \hat{F}_{\rho\alpha}(x) ,
			\hat{F}_{\sigma\beta}(x) \bigr\} \bigl| \Omega \bigr\rangle
		\, .
\end{equation}
Computing this calls for a dimensionally regulated coincidence limit of the field strength
correlator~(\ref{F corr dS}).
The tensor structures in this limit reduce to,
\begin{equation}
\bigl( \partial_\mu y \bigr)
	\xrightarrow{x' \to x}
	0 \, ,
\qquad
\bigl( \partial_\nu y \bigr)
	\xrightarrow{x' \to x}
	0 \, ,
\qquad
\bigl( \partial_\mu \partial'_\nu y \bigr)
	\xrightarrow{x' \to x}
	- 2 H^2 g_{\mu\nu}
	\, ,
\end{equation}
while the only relevant derivative of the propagator function is 
inferred from~(\ref{power series}),
\begin{equation}
\frac{\partial \mathcal{F}_\nu}{\partial y}
	\xrightarrow{x' \to x}
	- \frac{ H^{D-2} }{ (4\pi)^{\frac{D}{2} } } \frac{\Gamma(D\!-\!1)}{4 \, \Gamma\bigl( \frac{D+2}{2} \bigr)} \, .
\end{equation}
This results in a finite coincidence limit of the field strength correlator in~$D\!\to\!4$,
\begin{equation}
\bigl\langle \Omega \bigr| \hat{F}_{\mu\nu}(x) \hat{F}_{\rho\sigma}(x') \bigl| \Omega \bigr\rangle
	\xrightarrow{x' \to x}
		\frac{ H^{D} }{ (4\pi)^{\frac{D}{2} } } \frac{2 \, \Gamma(D\!-\!1)}{ \Gamma\bigl( \frac{D+2}{2} \bigr)}
		\times
		g_{\mu[\rho} g_{\sigma]\nu}
		\xrightarrow{D\to4} 
		\frac{ H^4 }{ 8 \pi^2 }
		\times
		g_{\mu[\rho} g_{\sigma]\nu}
		\, ,
\end{equation}
which implies a vanishing expectation value of the gauge-invariant energy-momentum tensor.
\begin{equation}
\bigl\langle \Omega \bigr| \hat{T}_{\mu\nu} \bigl| \Omega \bigr\rangle 
	= 0 \, .
\label{T expectation}
\end{equation}

\medskip
\noindent{\bf Gauge-fixed energy-momentum tensor.}
When defined as a variation of the gauge-fixed action, the classical energy momentum 
tensor,
\begin{equation}
T_{\mu\nu}^\star = \frac{-2}{\sqrt{-g}} \frac{\delta S_\star}{\delta g^{\mu\nu}}
	= T_{\mu\nu} + T_{\mu\nu}^{\rm gf}
	\, ,
\label{Tstar def}
\end{equation}
contains an additional part descending from the gauge-fixing term~(\ref{intro gauge}),
\begin{equation}
T_{\mu\nu}^{\rm gf} = 
	- \frac{2}{\xi} A_{(\mu} \nabla_{\nu)} \nabla^\rho A_\rho
	+ \frac{g_{\mu\nu}}{\xi} \biggl[ A_\rho \nabla^\rho \nabla^\sigma A_\sigma
		+ \frac{1}{2} \bigl( \nabla^\rho A_\rho \bigr)^2 \biggr]
		\, .
\label{Tgf def}
\end{equation}
Classically this piece vanishes on-shell~\cite{Glavan:2022a} as its every term contains at least one power
of the primary first-class constraint from~(\ref{first class constraints: Lagr}). 
When promoting this quantity to an operator it should be noted that
some terms contain longitudinal and temporal components of the vector potential.
This is why special attention needs to be given to operator ordering. 
Products containing constraints should not be Weyl-ordered, but rather written 
in a way that the non-Hermitian constraint operator~$\hat{K}(\vec{x})$, 
defined in sections~\ref{subsec: Quantized photon} and~\ref{subsec: Subsidiary condition},
should be on the right of the product, and its conjugate on the left of the product. 
This is conveniently accomplished by using the decomposition of Hermitian constraints in terms
of he non-Hermitian ones from~(\ref{nonhermitian split}),
\begin{subequations}
\begin{align}
\hat{T}_{00}^{\rm gf} ={}&
	- a^{2-D} \biggl[
		\hat{K}_2^\dag \hat{A}_{0}
		+ \hat{A}_{0} \hat{K}_2
		+ \bigl( \partial_k \hat{K}_1^\dag \bigr) \hat{A}_k
		+ \hat{A}_k \bigl( \partial_k \hat{K}_1 \bigr)
		+ \frac{\xi a^{4-D}}{2} \hat{\Psi}_1 \hat{\Psi}_1 
	\biggr]
	\, ,
\label{T00 gf}
\\
\hat{T}_{0i}^{\rm gf} ={}&
	- a^{2-D} \Bigl[  \bigl( \partial_i \hat{K}_1^\dag \bigr) \hat{A}_{0}
		+ \hat{A}_{0} \bigl( \partial_i \hat{K}_1 \bigr)
		+ \hat{K}_2^\dag \hat{A}_{i}
		+ \hat{A}_{i} \hat{K}_2
	\Bigr]
	\, ,
\label{T0i gf}
\\
\hat{T}_{ij}^{\rm gf} ={}&
	- a^{2-D} \biggl[
	2 \bigl( \partial_{(i} \hat{K}_1^\dag \bigr) \hat{A}_{j)}
	+ 2 \hat{A}_{(i} \bigl( \partial_{j)} \hat{K}_1 \bigr)
	+ \delta_{ij} \biggl(
		\hat{K}_2^\dag \hat{A}_0
		+ \hat{A}_0 \hat{K}_2
\nonumber \\
&	\hspace{1.5cm}
		- \bigl( \partial_k \hat{K}_1^\dag \bigr) \hat{A}_k
		- \hat{A}_k \bigl( \partial_k \hat{K}_1 \bigr)
		- \frac{ \xi a^{4-D} }{2} \hat{\Psi}_1 \hat{\Psi}_1 \biggr)
	\biggr]
	\, .
\label{Tij gf}
\end{align}
\end{subequations}
For the terms containing only products of the Hermitian constraint operators this ordering is immaterial
since the non-Hermitian constraint and its conjugate commute. From~(\ref{T00 gf})--(\ref{Tij gf}) then
immediately follows that,
\begin{equation}
\bigl\langle \Omega \bigr| \hat{T}_{\mu\nu}^{\rm gf} \bigl| \Omega \bigr\rangle
	= 0 \, ,
\label{Tmn: gf}
\end{equation}
for all physical  states satisfying subsidiary constraints~(\ref{position subsidiary}).
Thus, the expectation value of the energy momentum tensor
 for a physically
de Sitter invariant state vanishes, and it is immaterial which one of the two 
definitions,~(\ref{T def})
or~(\ref{Tstar def}), one uses to obtain this result.
This mirrors the property that the two definitions give the same answer on-shell.

In Refs.~\cite{BeltranJimenez:2008enx,BeltranJimenez:2009oao,BeltranJimenez:2011nm} 
the question of operator ordering of the gauge-fixed energy momentum
tensor operator was not addressed, and consequently the results reported there 
contradict the requirement~(\ref{Tmn: gf}).
Recently the same issue was tackled in~\cite{Zhang:2022csl},
but again without addressing the operator ordering. They concluded that it has to vanish, but only
after regularization and renormalization implemented by adiabatic subtraction. While we agree with
the conclusion, the rationale is quite different.
One either needs to consider operator ordering carefully,
 leading identically to~(\ref{Tmn: gf}), 
or one 
can obtain the answer~(\ref{Tmn: gf})
by Weyl-ordering operators 
and introducing compensating Faddeev-Popov ghost fields,
as we have done in a companion letter~\cite{GlavanProkopec:2022}.
The latter approach is consistent with other findings for 
the energy-momentum tensor 
expectation value, see~{\it e.g.}~\cite{Belokogne:2016dvd,BarrosoMancha:2020fay}.

\section{Discussion}
\label{sec: Discussion}

Photon propagators are essential building blocks for any quantum loop computation involving
massless vector fields. Here we are concerned with photon propagators in de Sitter space, 
which is a maximally symmetric idealization of more realistic slow-roll primordial inflation.
The expectation for maximally symmetric spaces, such as de Sitter, is that the general covariant gauge 
is conceptually and practically the simplest and most convenient one to use,~\footnote{There exists
a photon propagator in de Sitter in a non-covariant gauge due to Woodard~\cite{Woodard:2004ut},
that is simpler than any covariant gauge propagator. This suggests that covariant gauges 
need not be
the simplest choice in de Sitter.} on the account of two reasons:
\begin{itemize}
\item[(i)] 
it produces maximally symmetric two-point functions allowing for preservation of manifest
covariance at intermediate steps of the computation, and

\item[(ii)]
it contains a free gauge-fixing parameter defining a one-parameter family of covariant gauges,
that can be used to check computed observables that cannot depend on it.
\end{itemize}
Both of these expectations are of great utility in de Sitter space loop computations. The first
provides an organizational principle for computations that are notoriously more difficult than their 
flat space counterparts. The second is particularly useful for studying gauge dependence and
quantum observables in inflation, the understanding of which has still not reached maturity,
especially for perturbative quantum gravity~\cite{Miao:2017feh}.

In this work we have shown that the two commonly held expectations stated above are in general not
consistent with each other.
It is possible for the photon propagator to be de Sitter invariant only in the Landau gauge~$\xi\!\to\!0$,
in which case we reproduce the result due to Tsamis and Woodard~\cite{Tsamis:2006gj}. For all other values
of the gauge-fixing parameter it is not possible to maintain de Sitter symmetry of the propagator,
while simultaneously
satisfying the equations of motion~{\it and} subsidiary conditions. It is the latter that was not
checked for the covariant gauge photon propagators reported in the literature previously~\cite{Glavan:2022a},
and it is responsible for breaking of the de Sitter symmetry.
The de Sitter breaking part pertains to the pure gauge sector of the free theory, and in that
it has no physical content. However, when interactions are considered this is not 
as straightforward, and gauge-fixing has to be implemented correctly in order to guarantee
correct results for loop corrections.

Our main result is the photon two-point function in the general covariant gauge 
on the expanding Poincar\'{e} patch of de Sitter 
space,~\footnote{We do not consider here spatially compact global coordinates on de Sitter,
where the problem of linearization instability arises~\cite{Miao:2009hb}.}
that takes the following covariantized form,
\begin{equation}
i \bigl[ \tensor*[_\mu^{\scr - \!}]{\Delta}{_\nu^{\scr \! +}} \bigr] (x;x')
	= \bigl( \partial_\mu \partial'_\nu y \bigr) \, \mathcal{C}_1(y)
		+ \bigl( \partial_\mu y \bigr) \bigl( \partial'_\nu y \bigr) \, \mathcal{C}_2(y)
		+ \bigl( \partial_\mu u \bigr) \bigl( \partial'_\nu u \bigr)  \, \mathcal{C}_4 
		\, ,
\label{discussion propagator}
\end{equation}
The first two terms comprise the de Sitter invariant part, where both the tensor structures and the
scalar structure functions given in~(\ref{C1}) and~(\ref{C2}) are constructed from the 
de Sitter invariant distance~$y$ only.
This part was already obtained in previous 
works~\cite{Allen:1985wd,Tsamis:2006gj,Youssef:2010dw,Frob:2013qsa}.
The contribution of our analysis is the unexpected
third term, composed out of the de Sitter breaking tensor structure constructed from
variable~$u\!=\!\ln(aa')$,
multiplied by a constant,
\begin{equation}
\mathcal{C}_4 = \xi \times 
	\frac{H^{D-4}}{ (4\pi)^{\frac{D}{2}} } \frac{ \Gamma(D\!-\!1) }{ (D \!-\!1) \, \Gamma\bigl( \frac{D}{2} \bigr) }
	\, ,
\label{summary: C4}
\end{equation}
that vanishes only in the exact gauge limit~$\xi\!\to\!0$. This contribution is a homogeneous solution 
of the equation of motion, and would never be seen by considering a de Sitter invariant {\it Ansatz} for
the two-point function. It indeed is true that the de Sitter invariant part of~(\ref{discussion propagator})
satisfies the equation of motion independently, as was shown in~\cite{Allen:1985wd,Tsamis:2006gj,Youssef:2010dw,Frob:2013qsa}. However, the quantum subsidiary conditions  from 
Sec.~\ref{subsec: Quantized photon} and~\ref{subsec: Subsidiary condition}. 
are not satisfied unless the de Sitter breaking term is added.
These subsidiary
conditions are equivalent to the Ward-Takahashi identity for free two-point functions one obtains
from BRST quantization, that we considered in a companion letter~\cite{GlavanProkopec:2022}.

\bigskip

We derived our result~(\ref{discussion propagator}--\ref{summary: C4})
by starting from the first principles of canonical quantization.
This required us to first impose the general covariant gauge for the photon 
field as the multiplier
gauge in the classical theory in Sec.~\ref{sec: Photon in FLRW}, 
which treats all components of the vector field democratically, and
preserves general covariance. Such gauge fixing implies that the dynamics is given by the 
gauge-fixed action, and kinematics by the subsidiary first-class constraints.
The quantization, outlined in Sec.~\ref{sec: Field operator dynamics}, 
contains two aspects as well -- the gauge-fixed dynamics is quantized
by the usual rules of canonical quantization, while the constraints are implemented as conditions
on the space of states that physical states have to satisfy. In that respect the construction structurally
parallels Gupta-Bleuler quantization, but is formulated only in terms of the canonical structure
and is divorced from the symmetries of the theory. 

The symmetries of the quantum state are considered separately
from quantization in Sec.~\ref{sec: Choice of state}, where we construct
 the state by requiring it to be physically de Sitter invariant, meaning that
the expectation values of de Sitter generators descending from the original gauge invariant action
must all vanish. This fixes only the transverse part of the state that describes
the physical polarizations only. 
Then we construct the scalar sector of the state by requiring the full
state to be an eigenstate of the de Sitter symmetry generators of the gauge-fixed action. 
All of this was implemented in momentum space. 
Nevertheless, the integrals over all the modes
have to be performed to obtain the two-point function. 
When performed, integrals yield
 a de Sitter breaking two-point function in position space. 
While somewhat unintuitive, this is not the first
example of position-space two-point functions breaking de Sitter invariance that was imposed in 
momentum space. The best known example is the massless, minimally coupled scalar for which the CTBD
mode function is associated to the state that is an eigenstate of all the de Sitter symmetry generators,
but that fails to produce an infrared-finite two-point function, implying that it
does not exist.
For the massless photon we consider here de Sitter breaking is finite.

Our propagator in~(\ref{discussion propagator}--\ref{summary: C4})
and~(\ref{C1}--\ref{C2})
 is suitable for perturbative calculations in de Sitter space,
for Abelian and non-Abelian theories alike,
and can be also used to estimate the perturbative effects of interactions in general inflationary 
spacetimes.~\footnote{
 See Refs.~\cite{Prokopec:2007ak,Prokopec:2006ue,Prokopec:2008gw}
for how to tackle the problem of interactions beyond perturbation theory.
} Among the simplest such applications are the field strength correlator and the 
one-loop energy-momentum tensor discussed
in Sec.~\ref{sec: Simple observables}.

\bigskip

We expect that understanding of the gauge field quantization in inflation~\cite{Glavan:2022a},
and the construction of the two-point functions in de Sitter presented in this work 
and in the companion letter~\cite{GlavanProkopec:2022}, will allow us to 
tackle the questions faced when constructing
propagators in realistic inflationary spacetimes and when investigating the effects 
of interactions. Already in power-law inflation the situation is more complicated as there is
no enhanced symmetry, just the cosmological ones, and the construction of propagators is more 
involved~\cite{Janssen:2008px,Glavan:2020zne,Domazet:2023}.
The goal is to put such propagators to use and quantify how departures from exact de Sitter 
space influence the effects of quantum loop corrections. The understanding of 
this issue
 is important, but it
 is still in its early stages~\cite{Frob:2018tdd,Glavan:2019uni,Katuwal:2021kry}.

\section*{Acknowledgements}

We thank Ted Jacobson and Richard Woodard for discussions on the topic.
D.G. was supported by the Czech Science 
Foundation (GA\v{C}R) grant 20-28525S. 
This work is part of the Delta ITP consortium, a program of the Netherlands Organisation
for Scientific Research (NWO) that is funded by the Dutch Ministry of Education, Culture
and Science (OCW) --- NWO project number 24.001.027.

\appendix

\section{Noether currents and charges}
\label{app: Noether currents and charges}

Here we supply additional intermediate expressions for Sec.~\ref{sec: Choice of state}.

\subsection{Noether currents for gauge invariant action}

The global transformations from Sec.~\ref{subsec: De Sitter symmetries}
indeed are symmetries of the gauge invariant action~(\ref{action}). However, these are 
defined up to gauge transformations, which do not carry any information about
global de Sitter symmetries. Therefore, there are an ambiguities in the definition 
of global symmetries for the gauge-invariant action. However, 
there is a convenient way to fix these ambiguities
by requiring that the currents associated with the de Sitter symmetries be gauge-invariant
off-shell (they are invariant on-shell always, and currents are conserved on-shell only anyway).
To this end, it is more convenient to write the transformations of Sec.~\ref{subsec: De Sitter symmetries},
respectively, in the following form,
\begin{align}
A_\mu \rightarrow{}& 
	A_\mu - \alpha_i F_{i\mu} \, ,
\\
A_\mu \rightarrow{}&
	A_\mu + 2 \omega_{ij} x_i F_{j\mu} \, ,
\\
A_\mu \rightarrow{}&
	A_\mu + \frac{\alpha}{a} F_{0\mu} - \alpha H x_i F_{i\mu} \, ,
\\
A_\mu \rightarrow{}&
	A_\mu - \frac{\theta_i x_i}{a} F_{0\mu}
	+
	\biggl[ H \theta_j x_j x_i
	+ \frac{\theta_i}{2H} \Bigl( \frac{1}{a^2} \!-\! 1 \!-\! H^2 x_jx_j \Bigr) \biggr] F_{i\mu}
	\, .
\end{align}
The Noether currents associated with these transformations are now gauge invariant off-shell,
\begin{align}
(\mathscr{P}_i)^\mu 
	={}&
	\frac{ \partial \mathscr{L} }{ \partial(\partial_\mu A_\nu) }  \bigl( - F_{i \nu} \bigr)
	+ \delta_i^\mu \mathscr{L}
	\, ,
\\
(\mathscr{M}_{ij})^\mu 
	={}&
	\frac{ \partial \mathscr{L} }{ \partial(\partial_\mu A_\nu) } 
	\Bigl( 2 x_{[i} F_{j]\nu}\Bigr)
	+ \delta_{[i}^\mu x_{j]} \mathscr{L} 
	\, ,
\\
(\mathscr{Q})^\mu 
	={}&
	\frac{ \partial \mathscr{L} }{ \partial(\partial_\mu A_\nu) } 
		\Bigl( \frac{1}{a} F_{0\nu} - H x_i F_{i\nu} \Bigr)
	- \Bigl( \frac{ \delta_0^\mu }{a} 
		- \delta_i^\mu H x_i \Bigr) \mathscr{L}
	\, ,
\\
(\mathscr{K}_{i})^\mu 
	={}&
	\frac{ \partial \mathscr{L} }{ \partial(\partial_\mu A_\nu) } 
	\biggl[
		- \frac{ x_i }{ a } F_{0\nu}
		+ H x_i x_j F_{j\nu}
		+ \frac{ 1 }{ 2 H } \Bigl( \frac{1}{a^2} \!-\! 1 \!-\! H^2 x_j x_j \Bigr) F_{i\nu} 
		\biggr]
\nonumber \\
&
	+ \biggl[ 
		\delta_0^\mu \frac{ x_i }{a}
		- \delta_j^\mu H x_i x_j
		- \frac{ \delta_i^\mu }{ 2 H } \Bigl( \frac{1}{a^2} \!-\! 1 \!-\! H^2 x_j x_j \Bigr) \biggr]
			\mathscr{L}
			\, ,
\end{align}
where the gauge invariant Lagrangian and its derivative in expressions above are,
\begin{equation}
\mathscr{L}
	=
	- \frac{ \sqrt{-g} }{4} g^{\mu\rho} g^{\nu\sigma} F_{\mu\nu} F_{\rho\sigma}
		\, ,
\qquad \quad
\frac{ \partial \mathscr{L} }{ \partial(\partial_\mu A_\nu) } 
	=
	- \sqrt{-g} \, g^{\mu\rho} g^{\nu\sigma} F_{\rho\sigma}
		\, .
\label{Lagrangians}
\end{equation}
Whether the conserved currents are defined as off-shell gauge invariant or not, 
they lead to the same conserved charges, given in~(\ref{Pi})--(\ref{Ki}).

\subsection{Noether currents for gauge-fixed action}

The Noether currents associated with the global transformations from Sec.~\ref{subsec: De Sitter symmetries} are,
\begin{align}
(\mathscr{P}_i^\star)^\mu 
	={}&
	\frac{ \partial \mathscr{L}_\star}{ \partial(\partial_\mu A_\nu) } \bigl( - \partial_i A_\nu \bigr)
	+ \delta_i^\mu \mathscr{L}_\star
	\, ,
\\
(\mathscr{M}_{ij}^\star)^\mu 
	={}&
	\frac{ \partial \mathscr{L}_\star}{ \partial(\partial_\mu A_\nu) } 
	\Bigl( 2 x_{[i} \partial_{j]} A_\mu
		+ 2 \delta_{\mu [i} A_{j]} \Bigr)
	+ 2 \delta_{[i}^\mu  x_{j]} \mathscr{L}_\star
	\, ,
\\
(\mathscr{Q}^\star)^\mu 
	={}&
	\frac{\partial \mathscr{L}_\star}{ \partial ( \partial_\mu A_\nu ) }
		\biggl[ \frac{ 1 }{a} \partial_0 A_\nu
		- H x_i \partial_i A_\nu
		- H A_\nu \biggr]
	- \Bigl( \frac{ \delta^\mu_0 }{a} 
		- \delta^\mu_i H x_i \Bigr) \mathscr{L}_\star
	\, ,
\\
(\mathscr{K}_{i}^\star)^\mu 
	={}&
	\frac{\partial \mathscr{L}_\star}{ \partial ( \partial_\mu A_\nu ) }
	\biggl[ - \frac{ x_i }{ a } \partial_0 A_\nu
	+ H x_i x_j \partial_j A_\nu
		+ \frac{ 1 }{ 2 H } \Bigl( \frac{1}{a^2} \!-\! 1
			\!-\! H^2 x_j x_j \Bigr) \partial_i A_\nu 
\nonumber \\
&	\hspace{0.5cm}
	+ H x_i A_\nu
	- \frac{ 1 }{ a } \bigl( \delta_\nu^0 A_i
		 + \delta_\nu^i A_0 \bigr)
	+ H \bigl( \delta_\nu^i x_j A_j - \delta_\nu^j x_j A_i \bigr) 
	\biggr]
\nonumber \\
&	\hspace{1cm}
	+ \biggl[
		\delta_0^\mu \frac{ x_i }{a}
		- \delta_j^\mu H x_i x_j
		- \frac{ \delta_i^\mu }{2H} \Bigl( \frac{1}{a^2} \!-\! 1 \! -\! H^2 x_j x_j \Bigr) 
		\biggr]
		\mathscr{L}_\star
		\, ,
\end{align}
where the gauge fixed Lagrangian and its derivative in the expressions above are,
\begin{equation}
\mathscr{L}_\star
	=
	\mathscr{L}
	- \frac{\sqrt{-g} }{2\xi} \bigl( g^{\mu\nu} \nabla_\mu A_\nu \bigr)^{\!2}
		\, ,
\qquad
\frac{ \partial \mathscr{L}_\star}{ \partial(\partial_\mu A_\nu) } 
	=
	\frac{ \partial \mathscr{L}}{ \partial(\partial_\mu A_\nu) } 
	- \frac{\sqrt{-g}}{\xi} \bigl( g^{\rho\sigma} \nabla_\rho A_\sigma \bigr) g^{\mu\nu}
		\, .
\end{equation}
and where the gauge-invariant parts are already given in~(\ref{Lagrangians}).
The gauge-fixed conserved charges associated to the gauge-fixed conserved currents
are then given in~(\ref{Pi star})--(\ref{Ki star}).

\subsection{Scalar-transverse decomposition of charges}
\label{Scalar-transverse decomposition of charges}

Both the gauge invariant charges~(\ref{Pi})--(\ref{Ki})
 and the gauge fixed charges~(\ref{Pi star})--(\ref{Ki star})
reveal some of their structure when the fields they are comprised of are decomposed into the transverse and
longitudinal parts. In particular, it becomes clear how to order products of field operators when defining
quantum symmetry generators as observables in~(\ref{hatPi})--(\ref{hatKi}) 
and~(\ref{hatPi star})--(\ref{hatKi star}).
These decompositions are for spatial translations,
\begin{subequations}
\begin{align}
P_i ={}&
	P_i^{\scr T}
	+
	\int\! d^{D-1}x \, \Bigl( \Psi_2 A_i^{\scr T} \Bigr)
	\, ,
\label{Appendix A: Pi gi}
\\
P_i^\star ={}& 
	P_i^{\scr T}
	+
	\int\! d^{D-1}x\,
	\Bigl(
	\Psi_2 A_i^{\scr L} - \Psi_1 \partial_i A_0 \Bigr) 
	\, ,
\label{Appendix A: Pi gf}
\\
P_i^{\scr T} ={}&
	\int\! d^{D-1}x\,
	\Bigl( - \Pi_j^{\scr T} \partial_i A_j^{\scr T} \Bigr) 
	\, ,
\label{Appendix A: Pi T}%
\end{align}
\end{subequations}
for spatial rotations,
\begin{subequations}
\begin{align}
M_{ij} ={}&
	M_{ij}^{\scr T}
	+
	\int\! d^{D-1}x \, 
	\Bigl(
	2 x_{[i} A_{j]}^{\scr T} \Psi_2
	\Bigr)
	\, ,
\label{Appendix A: Mij gi}
\\
M_{ij}^{\star} ={}&
	M_{ij}^{\scr T}
	+
	\int\! d^{D-1}x\, 
	\Bigl(
	2 \Psi_1 x_{[i} \partial_{j]} A_0
	- 2 \Psi_2 x_{[i}  A_{j]}^{\scr L}
	\Bigr)
	\, ,
\label{Appendix A: Mij gf}
\\
M_{ij}^{\scr T} ={}&
	\int\! d^{D-1}x \, 
	\Bigl(
	2 x_{[i} F_{j]k}^{\scr T} \Pi_k^{\scr T}
	\Bigr)
	\, ,
\label{Appendix A: Mij T}
\end{align}
\end{subequations}
for dilations,
\begin{subequations}
\begin{align}
Q ={}& 
	Q^{\scr T}
	+
	\int\! d^{D-1}x \, \biggl[
		- \frac{ a^{3-D} }{2} \Psi_2 \nabla^{-2} \Psi_2
		- H \Psi_2 x_i A_i^{\scr T}
		\biggr]
		\, ,
\label{Appendix A: Q gi}
\\
Q^\star 
	={}& 
	Q^{\scr T}
	+
	\frac{1}{a}
	\int\! d^{D-1}x\, \biggl[ 
		- \frac{a^{4-D} }{2}\Psi_2 \nabla^{-2} \Psi_2
		- \frac{a^{4-D}}{2} \xi \Psi_1^2
		- \Psi_2 A_0 
\label{Appendix A: Q gf}
\nonumber \\
&	\hspace{3.5cm}
		+ \Psi_1 \partial_i A_i^{\scr L} 
		+ \mathcal{H} \Bigl( \Psi_2 x_i A_i^{\scr L}
		+ \partial_i\Psi_1 x_i  A_0  \Bigr) \biggr]
	\, ,
\\
Q^{\scr T}
	={}& 
	\int\! d^{D-1}x\, 
		\biggl[
		\frac{ a^{3-D} }{2}  \Pi_{i}^{\scr T} \Pi_{i}^{\scr T} 
		+ \frac{ a^{D-5} }{2}  (\partial_i A_j^{\scr T}) (\partial_i A_j^{\scr T}) 
		- H \Pi_{i}^{\scr T}  ( 1 \!+\! x_j \partial_j ) A_i^{\scr T} 
		\biggr]
		\, ,
\label{Appendix A: Q T}
\end{align}
\end{subequations}
and for spatial special conformal transformations,
\begin{subequations}
\begin{align}
K_i ={}&
	K_i^{\scr T}
	+
	\int \! d^{D-1}x \, 
	\biggl[
	\biggl( \frac{ a^{3-D} }{2} x_i \Psi_2 
	a^{3-D}\Pi_i^{\scr T}
	+ (D\!-\!3) H A_i^{\scr T} \biggr) \frac{ 1 }{\nabla^2} \Psi_2
\label{Appendix A: Ki gi}
\nonumber \\
&	\hspace{3.5cm}
	+ H x_i x_j  A_j^{\scr T} \Psi_2
	+ \frac{ 1 }{2H} \Bigl( \frac{1}{a^2} \!-\! 1 \!-\! H^2 x_j x_j \Bigr)
		A_i^{\scr T} \Psi_2
	\biggr]
	\, ,
\\
K_i^\star ={}& 
	K_i^{\scr T}
	+
	\int \! d^{D-1}x \, 
	\biggl[
		\biggl( 
		\frac{ a^{3-D} }{2} x_i \Psi_2 
		+ a^{3-D} \Pi_i^{\scr T}
		+ (D\!-\!3) H A_i^{\scr T}
		\biggr) \nabla^{-2} \Psi_2
		- \frac{1}{a} A_i^{\scr T}  \Psi_1
\nonumber \\
&
		+ \frac{a^{3-D} \xi}{2} x_i  \Psi_1 \Psi_1
		+ \frac{1}{a} x_i A_j^{\scr L} \partial_j \Psi_1
		+ \frac{1}{a} x_i A_0 \Psi_2
		+ (D\!-\!1) H x_i A_0 \Psi_1
		- H x_i x_j A_j^{\scr L} \Psi_2
\nonumber \\
&	
		+ H x_i x_j \partial_j A_0 \Psi_1
		+ \frac{1}{2H} \Bigl( \frac{1}{a^2} \!-\! 1 \!-\! H^2x_j x_j \Bigr) 
			\Bigl( \partial_i A_0 \Psi_1 - A_i^{\scr L} \Psi_2 \Bigr)
		\biggr] \, .
\label{Appendix A: Ki gf}
\\
K_i^{\scr T} ={}&
	\int \! d^{D-1}x \, 
	\biggl[
	- \frac{ x_i }{2a} \biggl(
		a^{4-D} \Pi_{j}^{\scr T} \Pi_{j}^{\scr T}
		+ a^{D-4} ( \partial_j A_k^{\scr T}) (\partial_j A_k^{\scr T})
		\biggr)
	+ H x_i x_j ( \partial_j A_k^{\scr T} ) \Pi_{k}^{\scr T}
\nonumber \\
&	\hspace{-1.1cm}
	+ \frac{ 1 }{2H} \Bigl( \frac{1}{a^2} \!-\! 1 \!-\! H^2 x_j x_j \Bigr) 
		\Pi_{k}^{\scr T} \partial_i A_k^{\scr T} 
	+ H x_j  A_j^{\scr T} \Pi_{i}^{\scr T}
	+ H x_i A_j^{\scr T} \Pi_{j}^{\scr T}
	- H x_j A_i^{\scr T} \Pi_{j}^{\scr T}
	\biggr]
	\, ,
\label{Appendix A: Ki T}
\end{align}
\end{subequations}
where the first-class constraints are~$\Psi_1 \!=\! \Pi_0$ and~$\Psi_2 \!=\! \partial_i \Pi_i^{\scr L}$.

\section{Identities for tensor structures}
\label{app: Identities for tensor structures}

Checking that the solution for the two-point function~(\ref{covariant basis})
satisfies the appropriate equation of motion and the subsidiary conditions from 
Sec.~\ref{subsec: Generalities} is facilitated by the following
covariant  identities for derivatives,
\begin{equation}
\bigl( \nabla_\mu \nabla_\nu y \bigr)
	= H^2 g_{\mu\nu} (2\!-\!y) \, ,
\qquad \qquad
\bigl( \nabla_\mu \nabla_\nu u \bigr)
	= - H^2 g_{\mu\nu} - \bigl( \partial_\mu u \bigr) \bigl( \partial'_\nu u \bigr) \, ,
\end{equation}
and tensor structure contractions,
\begin{subequations}
\begin{align}
g^{\mu\nu} \bigl( \partial_\mu y \bigr) \bigl(\partial_\nu y\bigr) 
	&= g'^{\rho\sigma} \bigl( \partial'_\rho y\bigr) \bigl(\partial'_\sigma y\bigr)
	= H^2 (4y \!-\! y^2) 
	\, ,
\\
g^{\mu\nu} \bigl( \partial_\mu y\bigr) \bigl(\partial_\nu \partial'_\rho y \bigr)
	&= H^2 (2\!-\!y) \bigl( \partial'_\rho y \bigr)
	\, , 
\\
g'^{\rho \sigma} \bigl(\partial_\mu \partial'_\rho y \bigr) \bigl( \partial'_\sigma y \bigr)
	&= H^2 (2\!-\!y) \bigl( \partial_\mu y \bigr)
	\, ,
\\
g^{\mu\nu} \bigl( \partial_\mu \partial'_\rho y \bigr) \bigl( \partial_\nu \partial'_\sigma y \bigr)
	&= 4 H^4 g'_{\rho \sigma} - H^2 \bigl(\partial'_\rho y \bigr) \bigl( \partial'_\sigma y \bigr)
	\, ,
\\
g'^{\rho\sigma} \bigl(\partial_\mu \partial'_\rho y \bigr) \bigl( \partial_\nu \partial'_\sigma y \bigr)
	&= 4 H^4 g_{\mu \nu} - H^2 \bigl(\partial_\mu y \bigr) \bigl(\partial_\nu y \bigr)
	\, .
\end{align}
\end{subequations}
These are applicable regardless of the~$i\varepsilon$ prescription in the distance functions,
except in one relevant case,
\begin{equation}
\bigl( \nabla_\mu \nabla_\nu 
y_{++} \bigr) \Bigl( \frac{y_{\scr ++}}{4} \Bigr)^{\! - \frac{D}{2}}
	=
	H^2 g_{\mu\nu} \bigl( 2 \!-\! y_{\scr ++} \bigr) \Bigl( \frac{y_{\scr ++}}{4} \Bigr)^{\! - \frac{D}{2}}
	+ \bigl( a^2 \delta_\mu^0 \delta_\nu^0 \bigr)
		\frac{ 4 (4\pi)^{\frac{D}{2}} }{ H^{D-2} \Gamma\bigl( \frac{D}{2} \bigr) }
			\frac{ i \delta^D(x\!-\!x') }{ \sqrt{-g} }
			\, ,
\end{equation}
that accounts for how the solution for the photon Feynman propagator produces local 
terms in~(\ref{Feynman EOM}) and~(\ref{Feynman subsidiary}).


\end{document}